\documentclass[11pt]{article}
\usepackage{comment}
\usepackage{lineno,hyperref}
\modulolinenumbers[5]
\usepackage{graphicx}
\usepackage{subcaption}
\usepackage[noadjust]{cite}
\usepackage{calc}
\usepackage{amssymb}
\usepackage{colortbl}
\usepackage{etoolbox}
\usepackage[table]{xcolor}
\usepackage{amsmath}
\usepackage{orcidlink}

\usepackage{paralist, tabularx}

\DeclareMathAlphabet{\mathpzc}{OT1}{pzc}{m}{it}

\newcommand{\hlm}{\mathpzc{h}_{\ell m}}

\def\lesssim{\mathrel{\hbox{\rlap{\hbox{\lower4pt\hbox{$\sim$}}}\hbox{$<$}}}}
\def\gtrsim{\mathrel{\hbox{\rlap{\hbox{\lower4pt\hbox{$\sim$}}}\hbox{$>$}}}}
\def\alt{\mathrel{\hbox{\rlap{\hbox{\lower4pt\hbox{$\sim$}}}\hbox{$<$}}}}
\def\agt{\mathrel{\hbox{\rlap{\hbox{\lower4pt\hbox{$\sim$}}}\hbox{$>$}}}}

\usepackage{mathtools}

\newenvironment{cititemize2}
{\begin{list}{$\bullet$}
        {\setlength{\topsep}{0pt}
         \setlength{\itemsep}{0pt}
         \setlength{\parsep}{0.25\parsep}
         \settowidth{\labelwidth}{$\bullet$}
         \setlength{\leftmargin}{1em}
}
}
{\end{list}}

\newcommand{\splitatcommas}[1]{%
  \begingroup
  \ifnum\mathcode`,="8000
  \else
    \begingroup\lccode`~=`, \lowercase{\endgroup
      \edef~{\mathchar\the\mathcode`, \penalty0 \noexpand\hspace{0pt plus 1em}}%
    }\mathcode`,="8000
  \fi
  #1%
  \endgroup
}

\def\gta{\ifmmode {\mathbin{\lower 3pt\hbox   
    {$\,\rlap{\raise 5pt\hbox{$\char'076$}}\mathchar"7218\,$}}}
    \else {${\mathbin{\lower 3pt\hbox
    {$\rlap{\raise 5pt\hbox{$\char'076$}}\mathchar"7218\,$}}}
    $}\fi}
\def\lta{\ifmmode {\,\mathbin{\lower 3pt\hbox   
    {$\,\rlap{\raise 5pt\hbox{$\char'074$}}\mathchar"7218\,$}}}
    \else {${\mathbin{\lower 3pt\hbox
    {$\rlap{\raise 5pt\hbox{$\char'074$}}\mathchar"7218\,$}}}
    $}\fi}

\newcommand{\beq}{\begin{equation}}
\newcommand{\eeq}{\end{equation}}
\newcommand{\bea}{\begin{eqnarray}}
\newcommand{\eea}{\end{eqnarray}}

\definecolor{darkperiwinkle}{RGB}{102, 102, 128}

\definecolor{light-gray}{gray}{0.9}

\usepackage[normalem]{ulem}
\usepackage{amsfonts} 

\newcommand{\blind}{0}
	
    \makeatletter
    \renewcommand\section{\@startsection {section}{1}{\z@}%
                                       {-3.5ex \@plus -1ex \@minus -.2ex}%
                                       {2.3ex \@plus.2ex}%
                                       {\normalfont\fontfamily{phv}\fontsize{16}{19}\bfseries}}
    \renewcommand\subsection{\@startsection{subsection}{2}{\z@}%
                                         {-3.25ex\@plus -1ex \@minus -.2ex}%
                                         {1.5ex \@plus .2ex}%
                                         {\normalfont\fontfamily{phv}\fontsize{14}{17}\bfseries}}
    \renewcommand\subsubsection{\@startsection{subsubsection}{3}{\z@}%
                                        {-3.25ex\@plus -1ex \@minus -.2ex}%
                                         {1.5ex \@plus .2ex}%
                                         {\normalfont\normalsize\fontfamily{phv}\fontsize{14}{17}\selectfont}}
    \makeatother
	
	\usepackage{amsmath}
	\usepackage{graphicx}
	\usepackage{enumerate}
	\usepackage{xcolor}
	\usepackage{url} 
	
\begin{document}

		\def\spacingset#1{\renewcommand{\baselinestretch}%
			{#1}\small\normalsize} \spacingset{1}
		
		\if0\blind
		{
			\title{\bf AI and extreme scale computing to learn and infer the physics of higher order gravitational wave modes of quasi-circular, spinning, non-precessing
			black hole mergers}
			\author{Asad Khan $^a$$^b$$^c$, E. A. Huerta $^a$$^b$$^d$, and Prayush Kumar $^e$\\
			$^a$ Data Science and Learning Division, Argonne National Laboratory, \\Lemont, Illinois 60439, USA \\
            $^b$ Department of Physics, University of Illinois at Urbana-Champaign, \\Urbana, Illinois 61801, USA \\
            $^c$ National Center for Supercomputing Applications, \\University of Illinois at Urbana-Champaign, Urbana, Illinois 61801, USA \\ 
            $^d$ Department of Computer Science, University of Chicago, \\Chicago, Illinois 60637, USA \\ 
            $^e$ International Centre for Theoretical Sciences, \\Tata Institute of Fundamental Research, Bangalore 560089, India}
			\maketitle
		} \fi
		
		\if1\blind
		{

            \title{Probabilistic AI to learn and infer the physics of higher order gravitational wave modes of quasi-circular, spinning, non-precessing black hole mergers}
			\author{Author information is purposely removed for double-blind review}
			
\bigskip
			\bigskip
			\bigskip
			\begin{center}
				{\LARGE\bf \emph{IISE Transactions} \LaTeX \ Template}
			\end{center}
			\medskip
		} \fi
		\bigskip


\begin{abstract}
\noindent We use artificial intelligence (AI) to learn 
and infer the physics of higher order 
gravitational wave 
modes of quasi-circular, spinning, 
non precessing binary black hole mergers. 
We trained AI 
models using 14 million waveforms, 
produced with the surrogate model \texttt{NRHybSur3dq8}, 
that include modes up to $\ell \leq 4$ and 
$(5,5)$, except for $(4,0)$ and $(4,1)$, 
that describe binaries with mass-ratios \(q\leq8\), 
individual spins $s^z_{\{1,2\}}\in[-0.8, 0.8]$, 
and inclination angle \(\theta\in[0,\pi]\).
Our probabilistic AI surrogates 
can accurately constrain the mass-ratio, 
individual spins, effective spin, 
and inclination angle of numerical relativity waveforms 
that describe such signal manifold. We compared the 
predictions of our AI models with Gaussian process regression, 
random forest, k-nearest neighbors, and linear regression, 
and with traditional 
Bayesian inference methods through the 
\texttt{PyCBC Inference} toolkit, finding that 
AI outperforms all these approaches in terms of accuracy, 
and are between three to four orders of 
magnitude faster than traditional Bayesian inference methods.
Our AI surrogates were trained 
within 3.4 hours using distributed training on 
1,536 NVIDIA V100 GPUs in the Summit supercomputer.
\end{abstract}

\maketitle


\section{Introduction}
\label{sec:intro}

\noindent The development of rigorous, 
reproducible, statistically and domain informed 
artificial intelligence (AI) is leading to 
remarkable breakthroughs in science and 
engineering ~\cite{DeepLearning}, and guiding 
human intuition to find new fundamental results in 
pure mathematics~\cite{ai_math}. AI 
applications in gravitational wave astrophysics are 
evolving from prototypes to production scale 
discovery frameworks. Since we developed the first AI 
models to create a scalable, computationally efficient 
method to search for and find 
gravitational waves~\cite{geodf:2017a,GEORGE201864}, 
this approach has been embraced and further developed 
by an international community of researchers~\cite{Nat_Rev_2019_Huerta,Huerta2020,cuoco_review}. 
To date, AI has been explored for a variety of signal 
processing tasks, including detection~\cite{2018GN,2020arXiv200914611S,Lin:2020aps,Wang:2019zaj,Fan:2018vgw,Li:2017chi,Deighan:2020gtp,Miller:2019jtp,Krastev:2019koe,2020PhRvD.102f3015S,Dreissigacker:2020xfr,Adam:2018prd,Dreissigacker:2019edy,2020PhRvD.101f4009B,2021arXiv210810715S,2021arXiv210603741S,2021arXiv210812430G}, 
gravitational wave denoising and data cleaning~\cite{shen2019denoising,Wei:2019zlc,PhysRevResearch.2.033066}, parameter estimation~\cite{Shen:2019vep,Gabbard:2019rde,Chua:2019wwt,Green:2020hst,Green:2020dnx,2021arXiv210612594D,2021arXiv211113139D}, 
rapid waveform production~\cite{Khan:2020fso,PhysRevLett.122.211101}, waveform forecasting~\cite{2021PhRvD.103l3023L,khan_huerta_zheng_forecast}, 
and early warning systems for 
multi-messenger sources~\cite{2020arXiv201203963W,Wei:2020sfz,2021PhRvD.104f2004Y}, including the modeling of multi-scale 
and multi-physics systems~\cite{2020PhRvD.101h4024R,2022arXiv220508663K,2022arXiv220312634R}, among others.

\begin{figure}[h!]
\centerline{
\includegraphics[width=\linewidth]{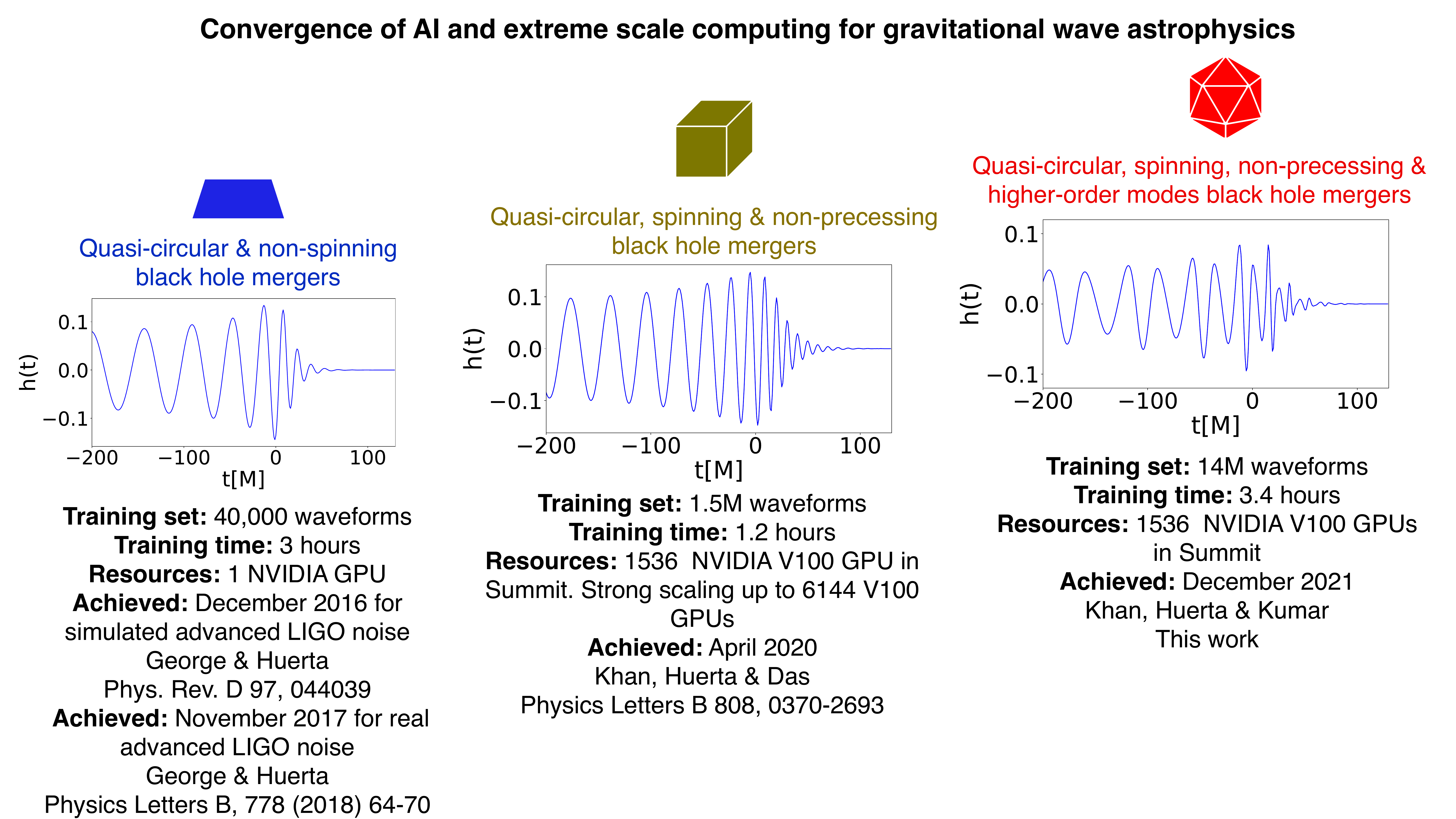}
}
\caption{\textbf{Convergence of AI and extreme scale computing} 
Progress harnessing AI and high performance computing to 
learn and describe the physics of gravitational wave 
signal manifolds. A signal manifold with richer complexity 
(from left to right) provides new opportunities to 
combine TB-size training datasets with extreme scale 
computing, allowing the development of novel 
distributed training algorithms 
and optimization schemes that incorporate physics and maths principles to accelerate the convergence and performance 
of AI surrogates.}
\label{fig:ai_hpc_convergence}
\end{figure}

Production scale AI frameworks for gravitational wave 
detection that harness high performance computing (HPC) and 
scientific data infrastructure have also 
been developed~\cite{huerta_nat_ast,Chaturvedi:2022suc}, furnishing 
evidence for the scalability, reproducibility and 
computational efficiency of AI-driven methodologies~\cite{2020arXiv200308394H}.
Figure~\ref{fig:ai_hpc_convergence} provides a glimpse 
of the rapid convergence of AI and extreme scale computing 
to study astrophysical scenarios that require waveforms 
with richer and more complex morphology. In view of 
these developments, it is time to further push the frontiers 
of AI applications to quantify their suitability to describe 
high-dimensional signal manifolds that contain waveforms 
whose morphology is significantly richer and much more 
complex than what has already been explored in the literature. 
This article represents a step in that direction. The 
driver we have selected for this study consists of 
characterizing higher order gravitational wave 
modes emitted by quasi-circular, spinning, non-precessing 
binary black hole mergers. We densely sample a 
parameter space that consists of binaries with mass-ratios 
\(q\leq 8\), individual spins $s^z_{\{1,2\}}\in[-0.8, 0.8]$, 
and inclination angle \(\theta\in[0,\pi]\) using a training dataset 
of 14M waveforms. The sheer size of 
this training dataset requires the combination of AI and 
HPC, and thus we harnessed the Summit supercomputer at 
Oak Ridge National Laboratory to 
reduce the training stage from months (using a single V100 GPU) 
to 3.4 hours using distributed training over 1536 NVIDIA V100 GPUs.

While this article showcases the convergence of AI and HPC 
for a computational grand challenge, the main goal of this 
analysis is to explore what new insights we may obtain by 
conducting AI-driven studies of the signal manifold of 
higher order gravitational wave modes. In particular, 
in this article we aim to explore the following 
open questions~\cite{KHAN2020135628}:

\begin{enumerate}
    \item Is it possible for AI to learn and 
    accurately characterize the physics of 
    high-dimensional gravitational wave signal manifolds? 
    \item Is it possible to exploit the computational efficiency 
    and scalability of AI to train models with tens of millions 
    of modeled waveforms and conduct fast data-driven 
    analyses with fully trained AI models?
    
    \item Is it true that probabilistic AI models provide more accurate inference predictions when we combine extreme scale computing to reduce time to insight, physics-informed AI architectures and optimization schemes to accelerate convergence, and high dimensional signal manifolds that include higher order wave modes to expose AI models to features and patterns that provide a detailed description of the waveform dynamics of black hole mergers?
    
    \item What insights do we gain when we characterize 
    gravitational wave signal manifolds with waveforms with 
    complex morphology?
\end{enumerate}

\noindent As we describe below, the answer to questions 1-3 
above is a resounding \texttt{YES}. In terms of new 
insights, we find that our deterministic and probabilistic 
AI models provide informative constraints for the 
mass-ratio, individual spins and inclination angle 
of higher order waveform modes. These are important results, 
since our previous work~\cite{KHAN2020135628} showed that, 
when we only consider \(\ell=|m|=2\) modes, it was 
difficult to constrain the individual spins of 
comparable mass-ratio systems, as well as the spin of 
the secondary for asymmetric mass-ratio systems. This 
study shows that the inclusion of higher order modes 
alleviates these problems, and provides 
an informed description of the 
ability of AI to characterize this high dimensional signal 
manifold. Throughout this paper we use geometric units in 
which \(G=c=1\).

This paper is organized as follows. Section~\ref{sec:method} 
describes the approach used to create our AI models. 
We present and discuss our findings in Section~\ref{sec:res}. 
Future directions of work are outlined in 
Section~\ref{sec:end}.

\section{Methods}
\label{sec:method}

\noindent Here we describe the datasets, AI architectures 
and training methods followed to create our AI models.

\begin{figure}[h!]
\centerline{
\includegraphics[width=0.33\linewidth]{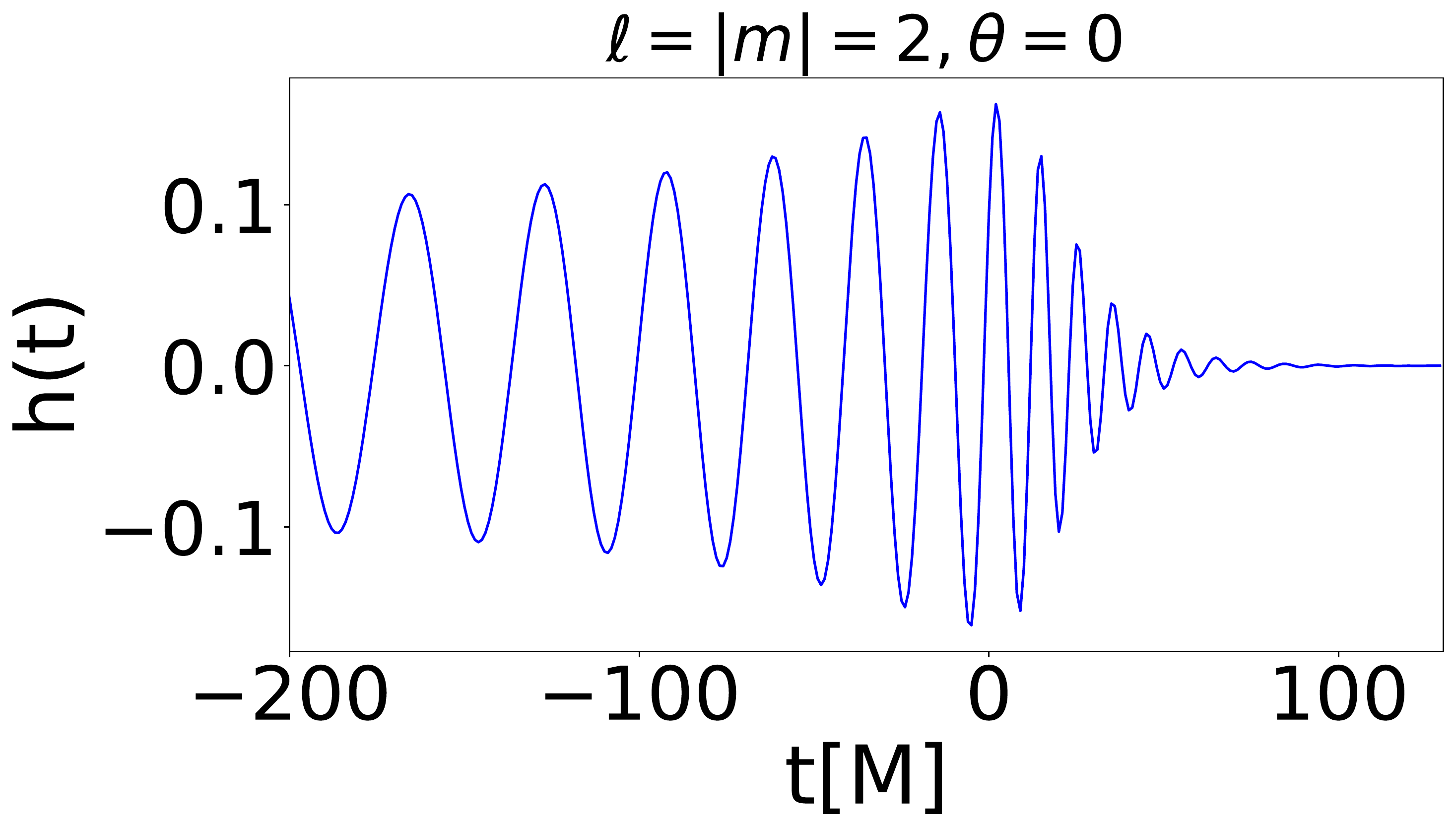}
}
\centerline{
\includegraphics[width=0.33\linewidth]{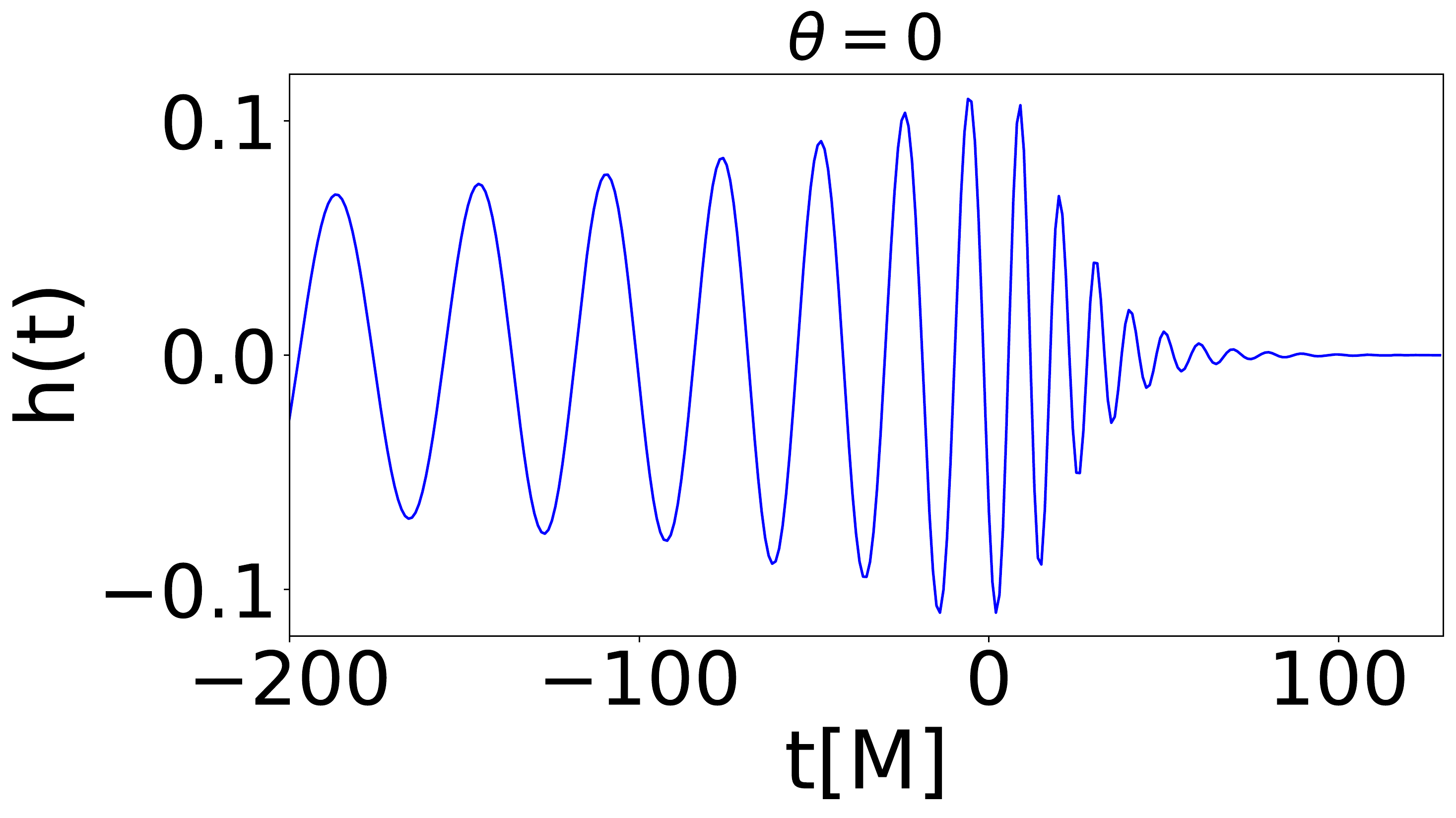}
\includegraphics[width=0.33\linewidth]{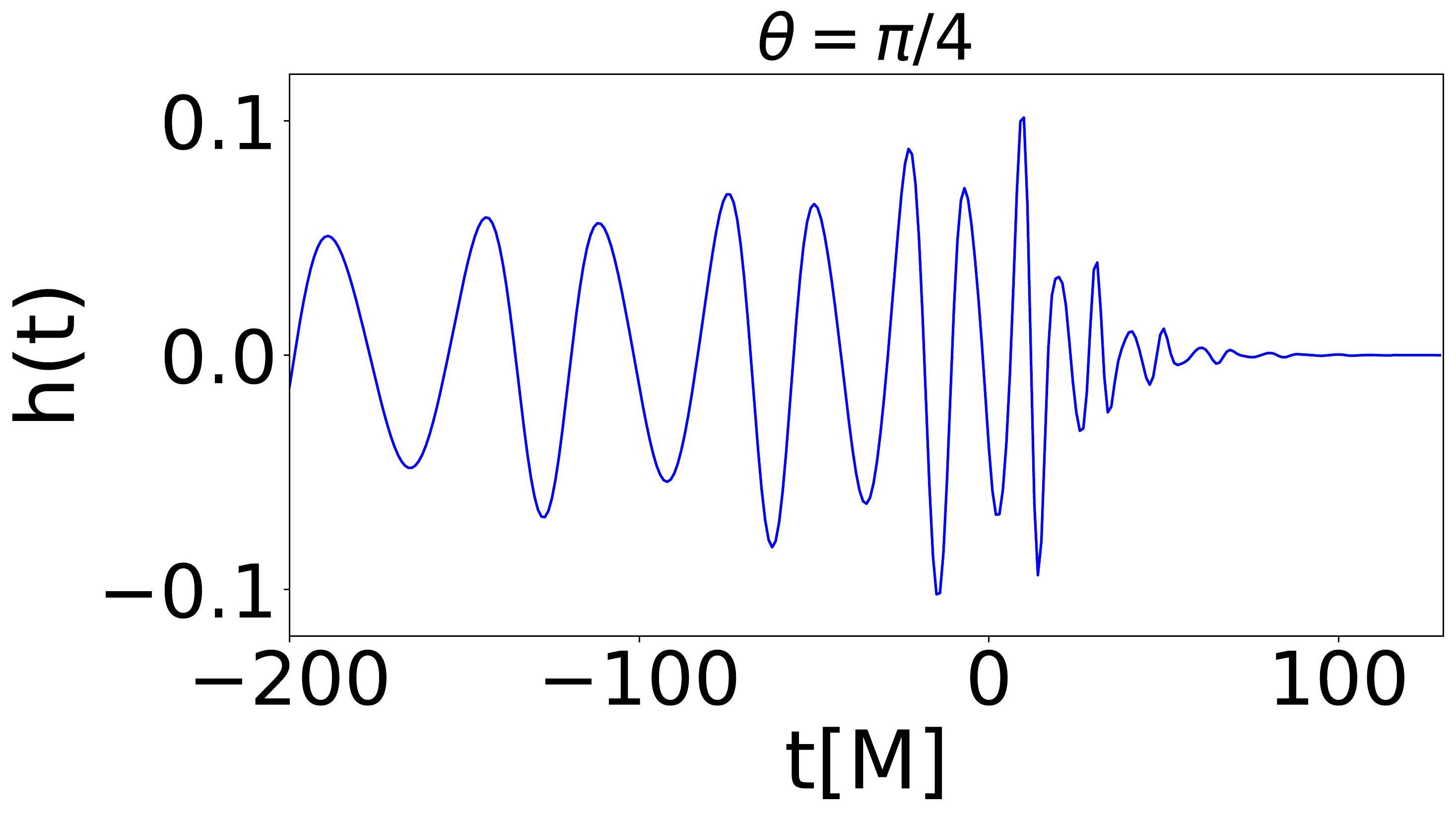}
\includegraphics[width=0.33\linewidth]{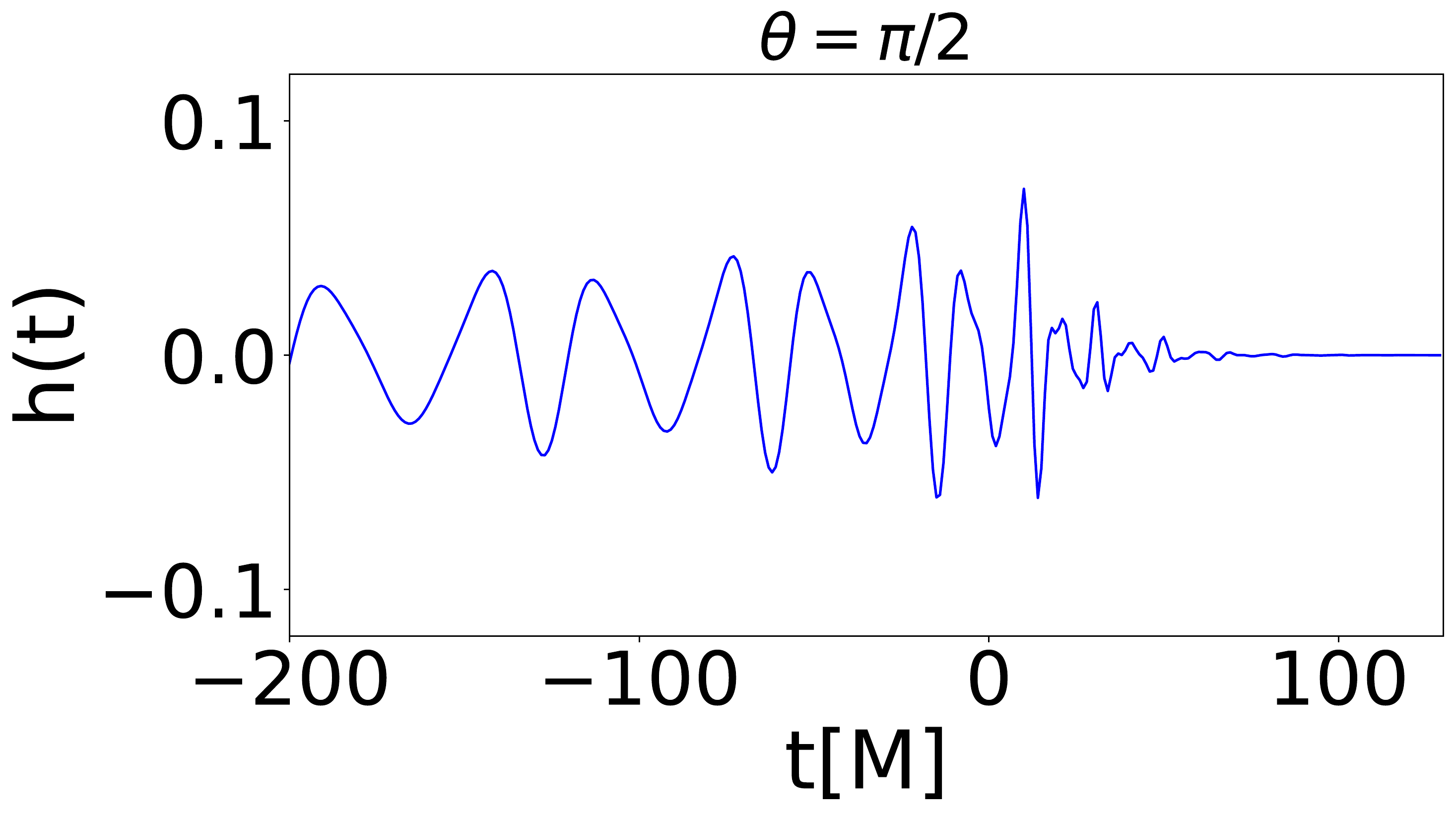}
}
\caption{\textbf{Impact of higher order modes}. 
For a binary black hole merger with parameters \(\{q,s_1^z, s_2^z\}=\{7.05, 0.77, 0.77\}\), we show waveforms 
with \((l=|m|=2; \theta=0)\) (top panel), and 
waveforms including higher order modes up to 
$\ell \leq 4$ and $(5,5)$, excluding the $(4,0)$ and 
$(4,1)$ modes, for \(\theta=\{0,\pi/4, \pi/2\}\) 
(bottom panels).}
\label{fig:22_vs_theta}
\end{figure}

\noindent \textbf{Datasets} We use the surrogate model \texttt{NRHybSur3dq8}~\cite{PhysRevD.99.064045} 
to generate time-series datasets that include 
both the plus, $h_{+}$, and cross, $h_{\times}$, polarizations. 
These may be represented as a 
complex time-series 
$h = h_{+} -ih_{\times}$. \(h\) may also be expressed 
as a sum of spin-weighted spherical harmonic 
modes, $\hlm$, on the 2-sphere~\cite{newman_penrose}

\begin{gather}
    h(t, \theta, \phi_0) = \sum^{\infty}_{\ell=2} \sum_{m=-l}^{l}
        \hlm(t) ~^{-2}Y_{\ell m}(\theta, \phi_0),
\label{eq:spherical_harm}
\end{gather}

\noindent where $^{-2}Y_{\ell m}$ are the spin-weight $\!\!-2$ 
spherical harmonics, $\theta$ is the inclination angle between the 
orbital angular momentum of the binary and line of sight to the detector, and $\phi_0$ is the initial binary
phase, that we set to zero in this study. Our waveforms 
include higher order modes with $\ell \leq 4$ and 
$(\ell, m) = (5,5)$, excluding the $(\ell, m)=(4,0)$ and $(4,1)$ modes; 
cover the time span 
\(t\in[-10,000\,\textrm{M}, 130\,\textrm{M}]\) with the 
merger peak occurring at \(t=0M\); and are sampled with a
time step \(\Delta t = 1\,\textrm{M}\). 
Figure~\ref{fig:22_vs_theta} presents a 
sample of waveforms that 
help visualize the importance of including higher order modes in 
terms of the amplitude and phase evolution, as well as the 
morphology of the ringdown phase. The top panel presents a waveform 
signal that only includes the \(\ell=|m|=2\) mode at an optimally 
oriented configuration, \(\theta=0\), that maximizes the amplitude 
of the signal for detectability purposes. 
The bottom panels show what new information may be obtained as 
we construct waveform signals that include higher order modes, 
i.e., the amplitude and phase of the pre-merger waveform 
signal exhibits novel, non-linear features as well as a richer 
and much more complex  ringdown phase. In stark contrast, the 
top panel will only change the amplitude of the waveform 
signal for angles \(\theta=\{\pi/4, \pi/2\}\), 
since the \(\ell=|m|=2\) mode signal is modulated by a 
constant multiplicative factor that is set by the inclination angle~\cite{1967JMP.....8.2155G,Blanchet:2013haa}.

\noindent \textbf{Training dataset} It consists 
of $\sim14$ million waveforms that cover a 
4-D parameter space that encompasses mass-ratio, individual spins 
and inclination angle \(\{q, s_1^z, s_2^z, \theta\}\), 
respectively. We generate it by sampling the mass-ratio $q\in[1,8]$ 
in steps of \(\Delta q = 0.1\); individual spins  
$s^z_i\in[-0.8, 0.8]$ in steps of \(\Delta s^z_i = 0.02\); 
and the inclination angle $\theta\in[0, \pi]$ in 
steps of \(\Delta \theta = 0.1\). 

\noindent \textbf{Validation and test datasets} Each of these 
sets consist of $\sim 800,000$ waveforms, and 
are generated by alternately sampling values 
that are inbetween the training set values.

\noindent \textbf{AI architecture} We use a slightly modified \texttt{WaveNet} \cite{oord2016wavenet} 
neural network architecture for our model. \texttt{WaveNet}'s 
main features that are relevant for this work include dilated 
causal convolutions,  gated  activation  units, and the 
usage  of residual and skip  connections. These features help 
capture long range correlations in the input time-series, and 
facilitate the training of deeper neural networks. 
Furthermore, since we are interested in regression analyses, 
we turn off the causal padding 
in the convolutional layers. We use a filter size of 2 in 
all convolutional layers and stack 3 residual blocks 
each consisting of 14 dilated convolutions. For a more in depth
discussion of the architecture we refer the reader to \cite{KHAN2020135628,oord2016wavenet}. The output from
the \texttt{WaveNet} is then fed into three separate branches 
of fully-connected layers. Each branch is trained to predict the 
mass ratio, \(q\), the effective spin parameters, \((S_{\textrm{eff}}\), \(\sigma_{\textrm{eff}})\), 
and the inclination angle, \(\theta\), respectively, where

\begin{equation}
    S_{\textrm{eff}} = \sigma_1 s^z_1 + \sigma_2 s^z_2\,, \quad \textrm{with}\quad \sigma_1 \equiv 1 + \frac{3}{4q} \quad \textrm{and}\quad \sigma_2 \equiv 1 + \frac{3q}{4}\,,
    \label{eq:eff_spin}
\end{equation}

\begin{equation}
   \textrm{and}\quad  \sigma_{\textrm{eff}} = \frac{m_1s^z_1 + m_2 s^z_2}{m_1 + m_2} = \frac{q s^z_1 + s^z_2}{1 + q}\,.
    \label{eq:alt_eff_spin}
\end{equation}

\noindent Since our goal is to predict $\{q, s^z_1, s^z_2, \theta\}$, 
we solve Eqs.~\eqref{eq:eff_spin} and~\eqref{eq:alt_eff_spin} 
in conjunction with the predicted $q$ values in order to extract 
the individual spins $s^z_i$.

\noindent \textbf{Training methodology} We employ mean-squared 
error (MSE) between the predicted and the ground-truth values
as the loss function. During training, we monitor the loss 
on the validation set to dynamically 
reduce learning rate as well as to stop training before 
over-fitting. We reduce the
learning rate by a factor of $2$ whenever the validation 
loss does not decrease for $3$ consecutive 
epochs, and stop the training when the validation loss 
does not decrease for $5$ consecutive epochs. 
Training the model on $256$ nodes, equivalent to 
1,536 NVIDIA V100 GPUs, in the
Summit Supercomputer then takes about $71$ epochs, using
\texttt{LAMB}~\cite{you2019large} optimizer with 
initial learning rate set to $0.001$.

\begin{figure}[h!]
\centerline{
\includegraphics[width=.4\linewidth]{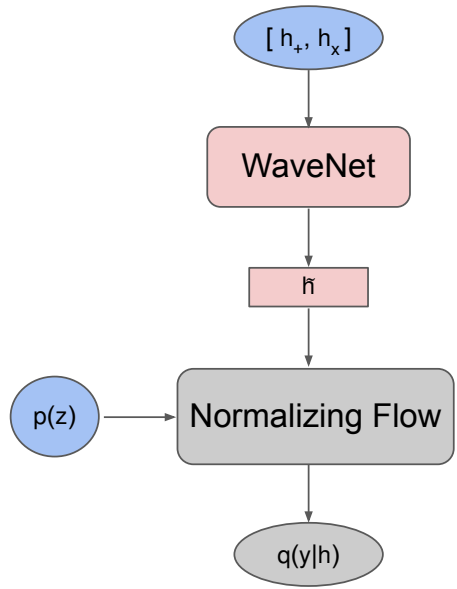}
}
\caption{\textbf{Normalizing flow model architecture}. Schematic 
representation that indicates how we feed the waveform 
$[h_+,h_x]$ into a \texttt{WaveNet} model to extract 
the feature vector $\tilde{h}$. Then the normalizing flow 
$f_{\tilde{h},\theta}$ takes in this feature vector 
and transforms the base distribution 
$\mathcal{N}(\mu=0, \Sigma=\mathbf{I})$ into the 
predicted 
posterior distribution $q(y|h)$.}
\label{fig:nflow}
\end{figure}

\noindent \textbf{Normalizing Flow:} In addition to the above 
analysis for point estimates, we also trained a normalizing flow 
model to estimate the posterior distribution. To do that, instead
of extracting the parameters directly, we use the \texttt{WaveNet} 
model as a feature extractor and then condition a normalizing flow 
model on the extracted features to estimate the posterior 
distribution. This method was first delineated in~\cite{green2021complete}, 
and then 
used in other studies~\cite{Shen:2019vep,2021arXiv210612594D}. 
We follow the same procedure, making use of the \texttt{nflows} library~\cite{nflows}.
Normalizing flow is an example of a ``likelihood-free''
inference method, and is made up of a composition of invertible maps 
to transform a simple base probability distribution (e.g., a multivariate Gaussian) 
into a desired posterior distribution which could be very 
complicated. The transformed distribution is then given by the 
change of variable formula:

\begin{equation}
    p_X(x) = p_Z(z) \left|\det [\mathbf{J}_f(z)]\right|^{-1}\,,
    \label{eq:nf}
\end{equation}

\noindent where $Z$ is the random variable for the base 
distribution, $X$ is the random variable for the transformed 
distribution, and $f$ is the normalizing flow (i.e., the 
invertible transformation), such that $X = f(Z)$. In our case, 
the goal is to model the 
conditional posterior distribution $p(y|h)$ for the parameters $y$ 
corresponding to the waveform strain $h$. We do this in two steps 
(as illustrated in Figure~\ref{fig:nflow}); 
first we pass the strain $h$ through the \texttt{WaveNet} model to extract a 
feature vector $\tilde{h}$. We then use a conditional version of 
normalizing flow $f_{\tilde{h},\vartheta}$, 
specifically a \texttt{neural spline flow}~\cite{durkan2019neural},
to transform the base Standard Multivariate Gaussian 
$\mathcal{N}(\mu=0, \Sigma=\mathbf{I})$ to the predicted 
posterior distribution 
$q(y|h)$. The function $f_{\tilde{h},\vartheta}$ therefore 
depends on the input 
waveform  $h$ (through the feature vector $\tilde{h}$), and is 
parameterized by 
learnable weights $\vartheta$. The normalizing flow model is then 
trained by 
updating the parameter $\vartheta$ so that predicted distribution 
$q(y|h)$ matches
the true posterior distribution $p(y|h)$. This is achieved 
by minimizing the 
negative log-likelihood, i.e., given a batch of $N$ ground-truth 
parameters $y_i$
and their corresponding waveforms $h_i$, the objective for the 
normalizing flow model is to minimize:

\begin{align}
    L = -\frac{1}{N} \sum_{i=1}^N \log q(y_i | h_i)
\end{align}

\noindent where, according to Equation~\ref{eq:nf}; 

\begin{align}
    q(y_i | h_i) = p_{\mathcal{N}}(z) \left|\det [\mathbf{J}_{f_{\tilde{h},\theta}}(z)]\right|^{-1}\,.
\end{align}

\section{Results}
\label{sec:res}

\noindent We present results using deterministic and probabilistic 
AI models, which we described in Section~\ref{sec:method}. 
As described above, these AI models have been designed to 
take in time-series waveform signals that includes both polarizations 
\((h_+, h_\times)\), and then output the most likely 
values for \(\{q, s_1^z, s_2^z,\theta\}\) that best 
reproduce the input signal. 

\subsection{Deterministic AI models} 
\begin{figure}[h!]
\centerline{
\includegraphics[width=0.33\linewidth]{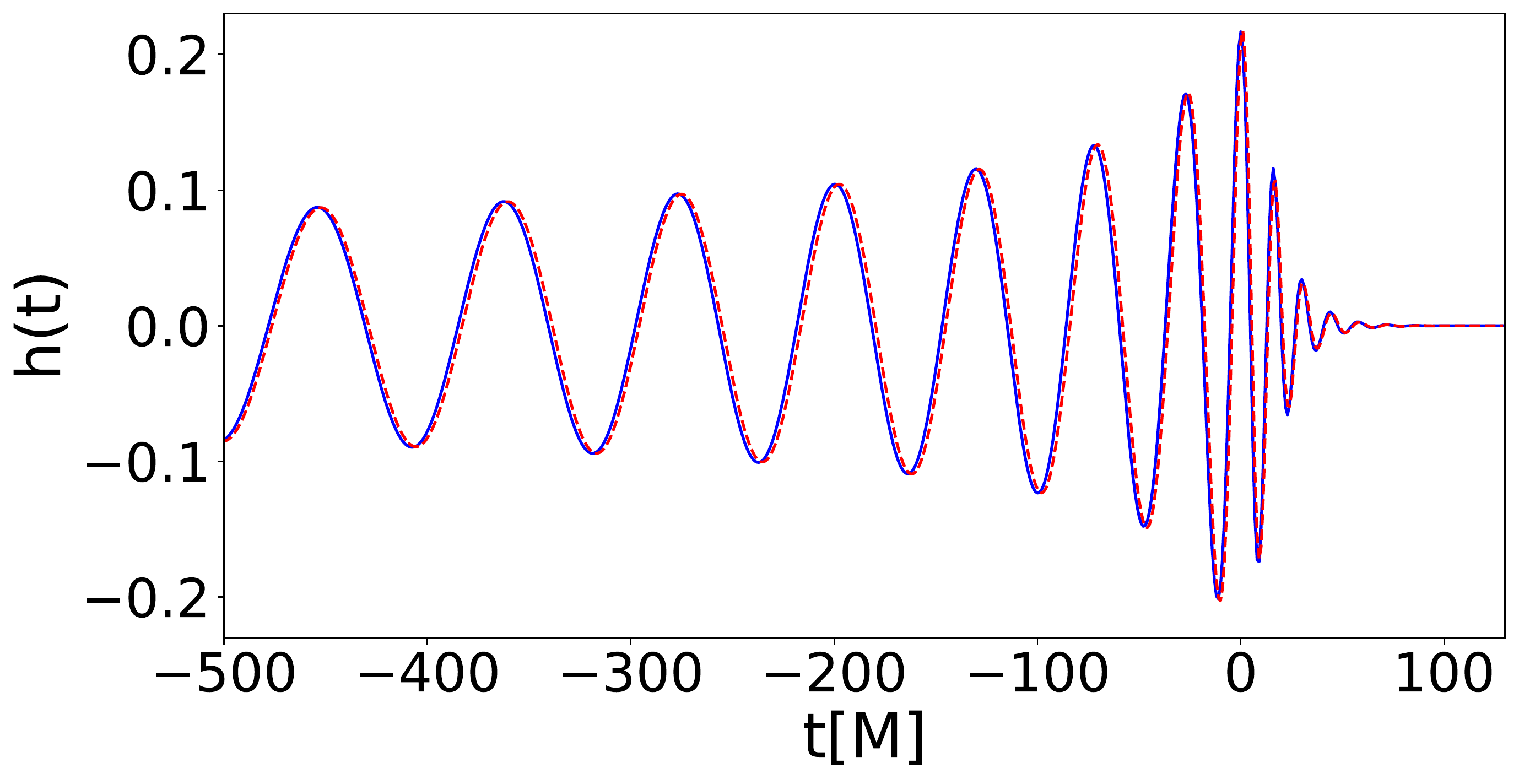}
\includegraphics[width=0.33\linewidth]{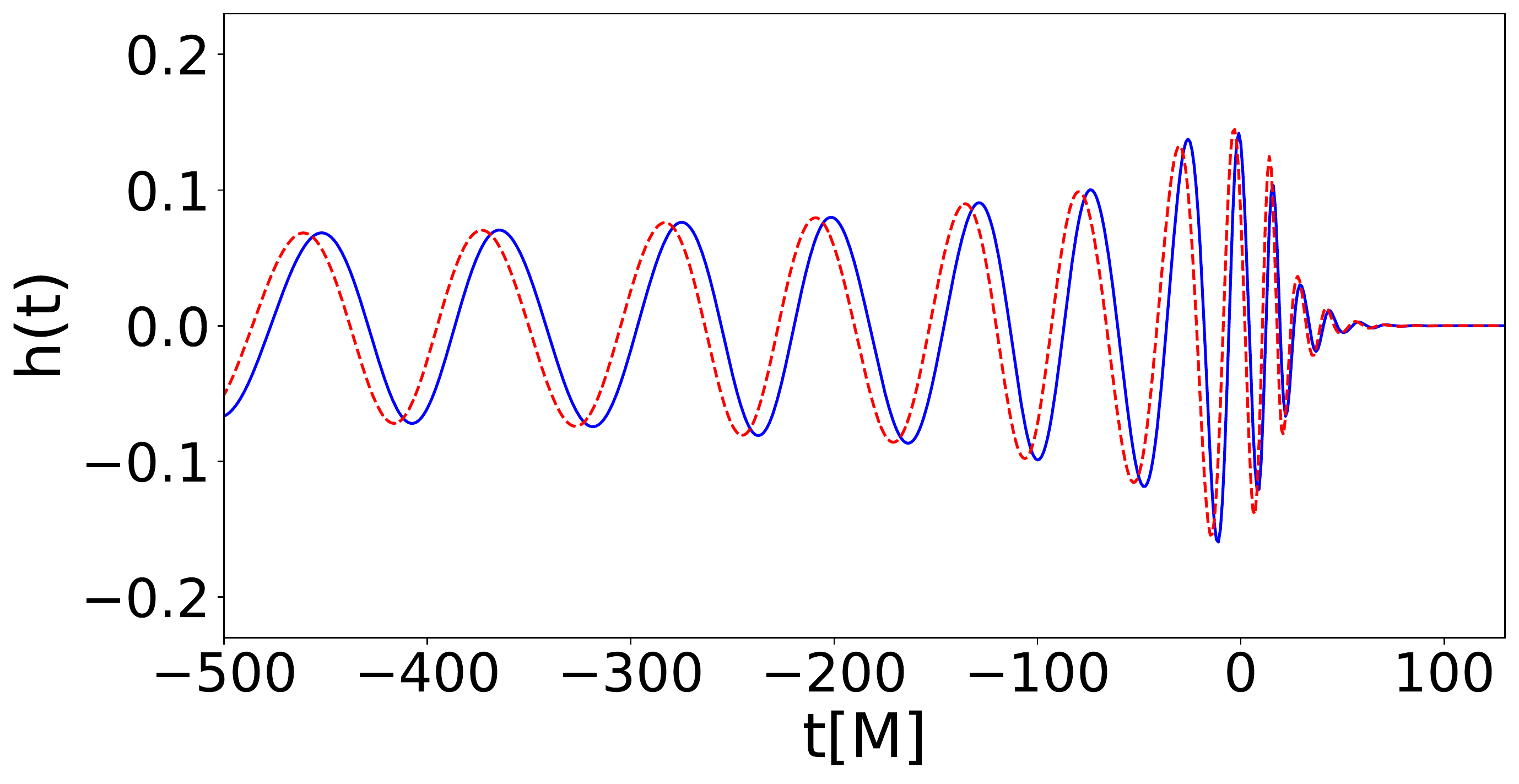}
\includegraphics[width=0.33\linewidth]{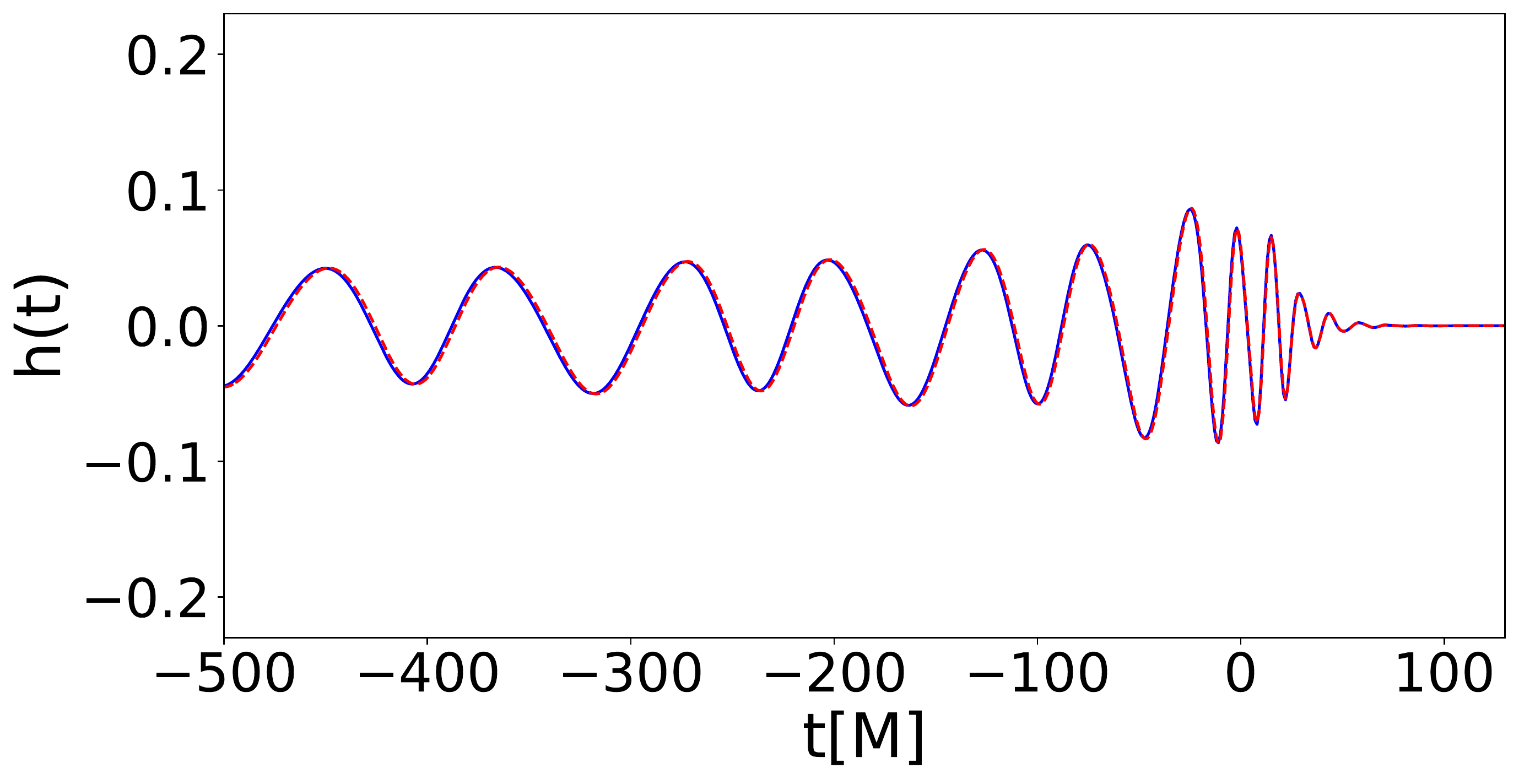}
} 
\centerline{
\includegraphics[width=0.33\linewidth]{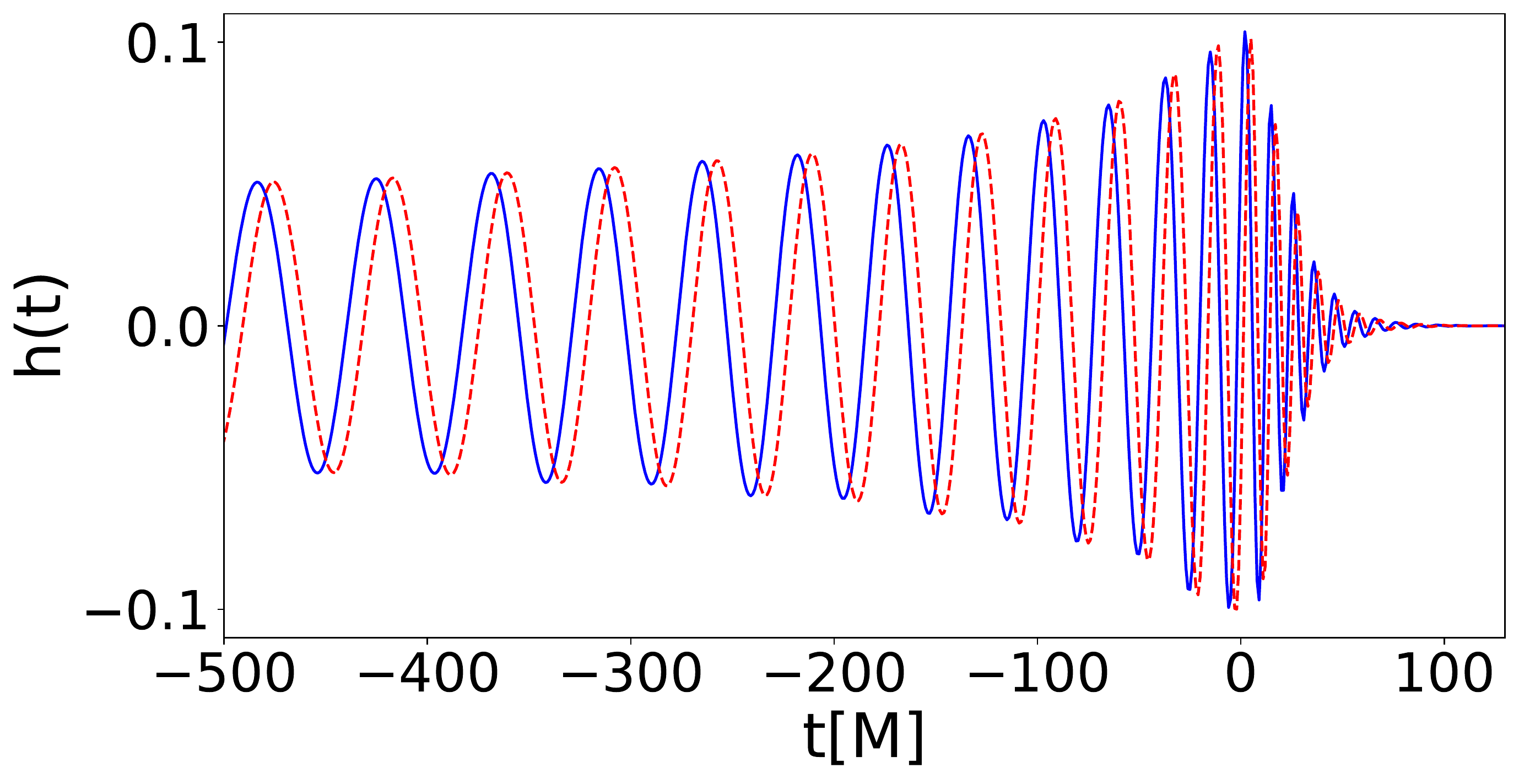}
\includegraphics[width=0.33\linewidth]{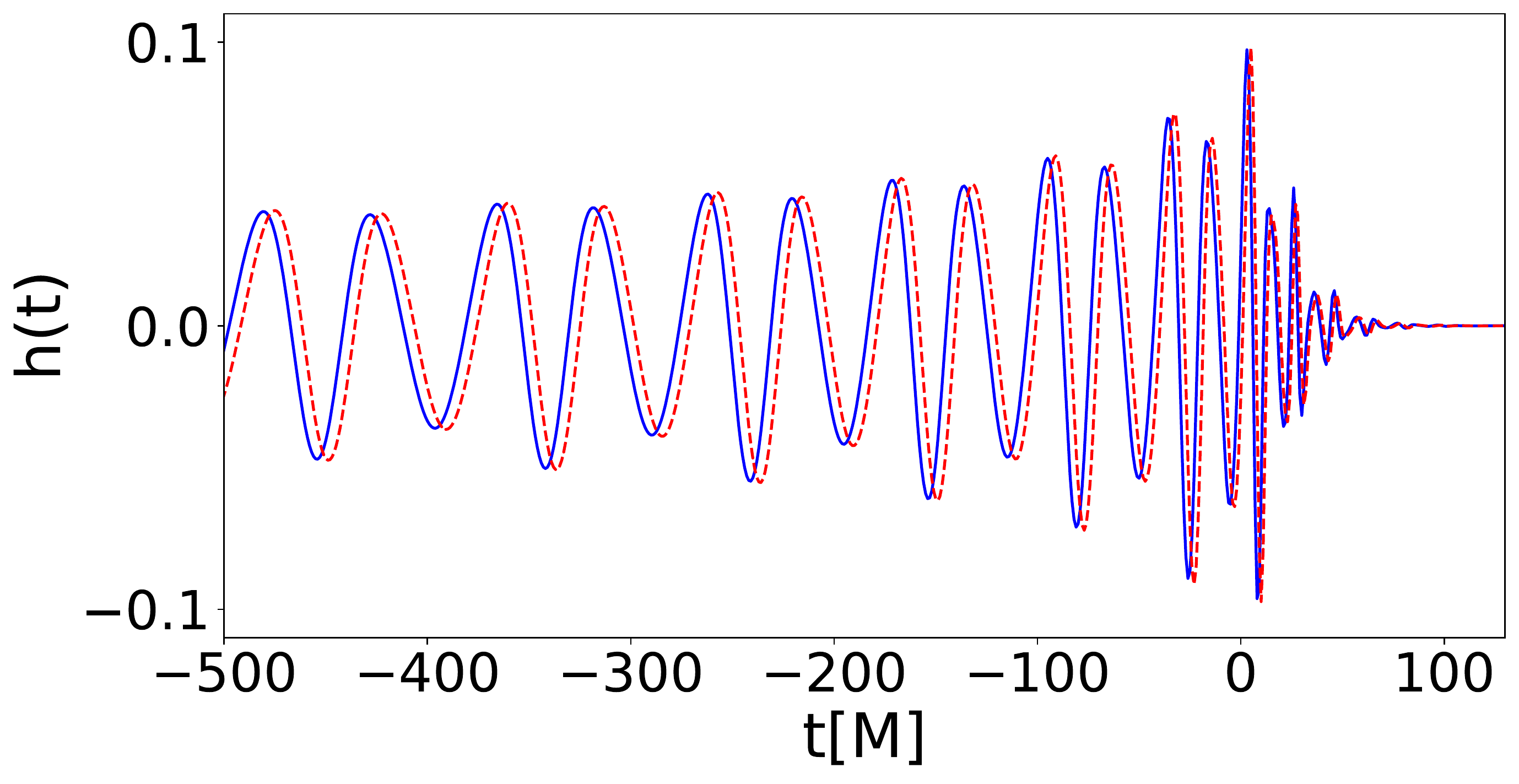}
\includegraphics[width=0.33\linewidth]{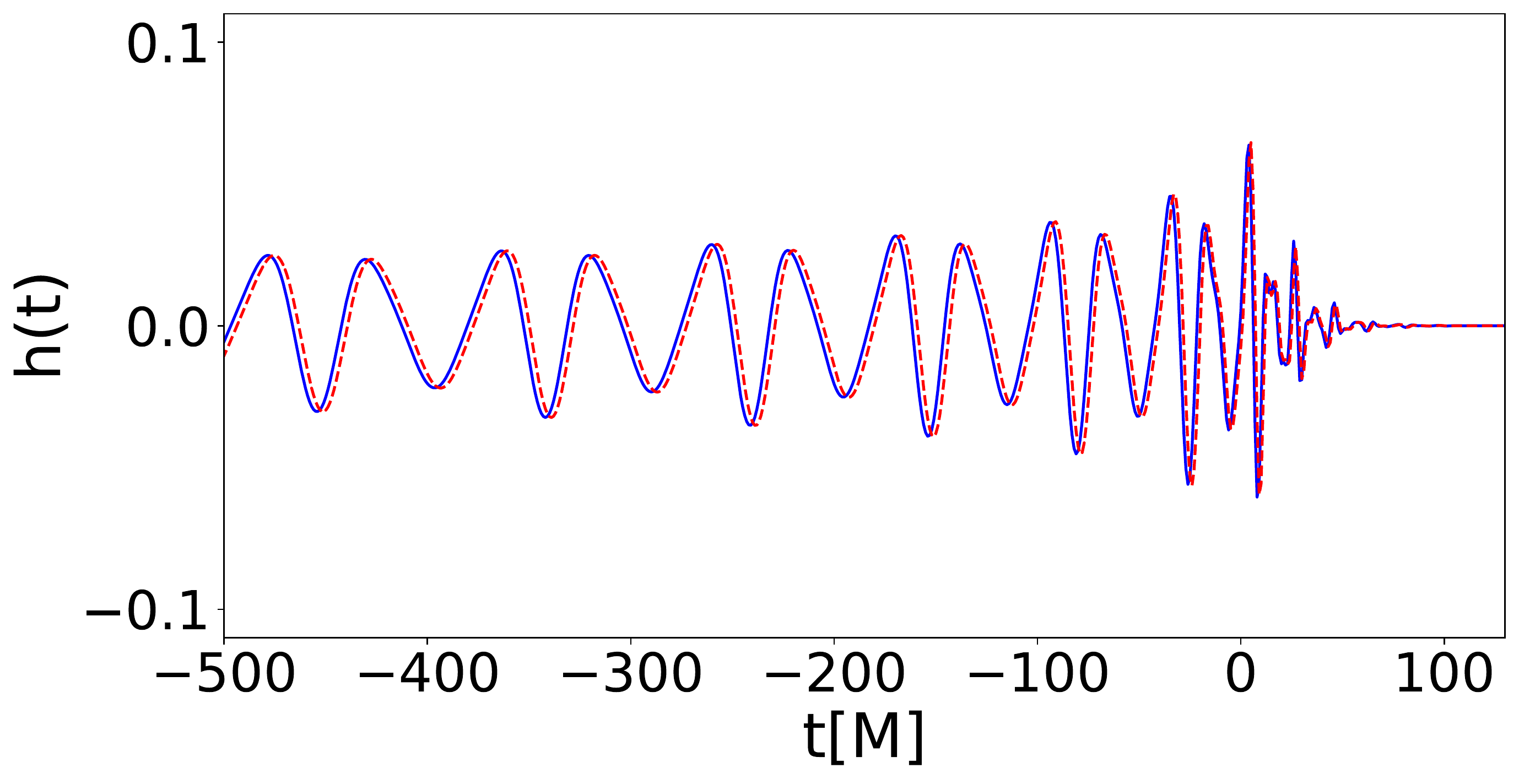}
}
\caption{\textbf{Comparison between ground truth and AI-predicted waveforms} Ground truth waveforms (in blue) and AI predicted 
waveforms (dotted red). Top panels: \(\{q, s_1^z, s_2^z\} = \{2.05, -0.79, 0.77\}\), bottom panels: \(\{q, s_1^z, s_2^z\} = \{7.85, 0.77, -0.79\}\).
We show waveforms, from left to right, with inclination angles \(\theta=\{0, \pi/4, \pi/2\}\). In all these cases, the overlap between ground 
truth and predicted waveforms is \({\cal{O}} \geq 0.98\).}
\label{fig:q_2_3}
\end{figure}

\noindent In Figure~\ref{fig:q_2_3} we provide a sample of results 
of the 
predictive capabilities of our deterministic AI models for a 
variety of input waveform signals. Note that 
ground truth waveforms are shown in 
blue, whereas waveforms whose parameters, 
\(\{q, s_1^z, s_2^z,\theta\}\), are 
predicted by AI are shown in dotted red. We quantify the 
accuracy of our AI models by computing the overlap, 
\({\cal{O}}(h_t|h_p)\),
between ground-truth, \(h_t\), and AI-predicted, 
\(h_p\), waveforms using 

\begin{align}
\label{over}
{\cal{O}}  (h_t|h_p)= \underset{ [t_c\, \phi_c]}{\mathrm{max}}\left(\hat{h}_t|\hat{h}_p{[t_c,\,  \phi_c]}\right)\,,\quad{\rm with}\quad \hat{h}_t=h_t\,\left(h_t | h_t\right)^{-1/2}\,,
\end{align}

\noindent where \({[t_c,\,  \phi_c]}\) are the amounts by
which the normalized waveform \(\hat{h}_p\) has been
time- and phase-shifted. 

\begin{figure}[h!]
\centerline{
\includegraphics[width=0.8\linewidth]{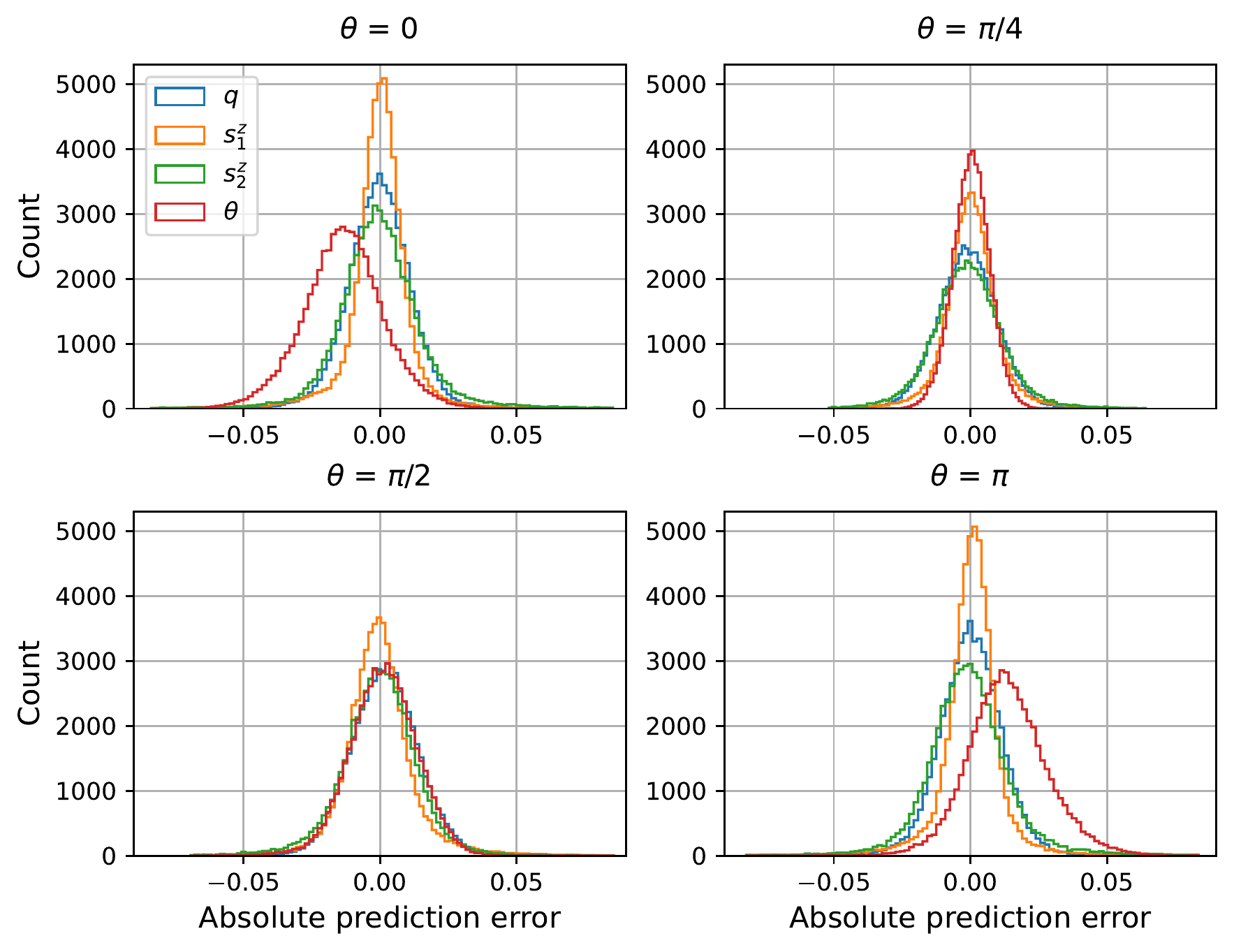}
} 
\caption{\textbf{Statistical results of deterministic AI models}
Absolute errors for all mass ratios, \(q\), and 
individual spins, \((s_1^z, s_2^z)\), for 
a sample of inclination angles. The true values of inclination angles
are $[0.054, 0.70, 1.57, 3.09]$ and are rounded-off for labeling above.}
\label{fig:ae_vs_theta}
\end{figure}

\begin{figure}[h!]
\centerline{
\includegraphics[width=0.8\linewidth]{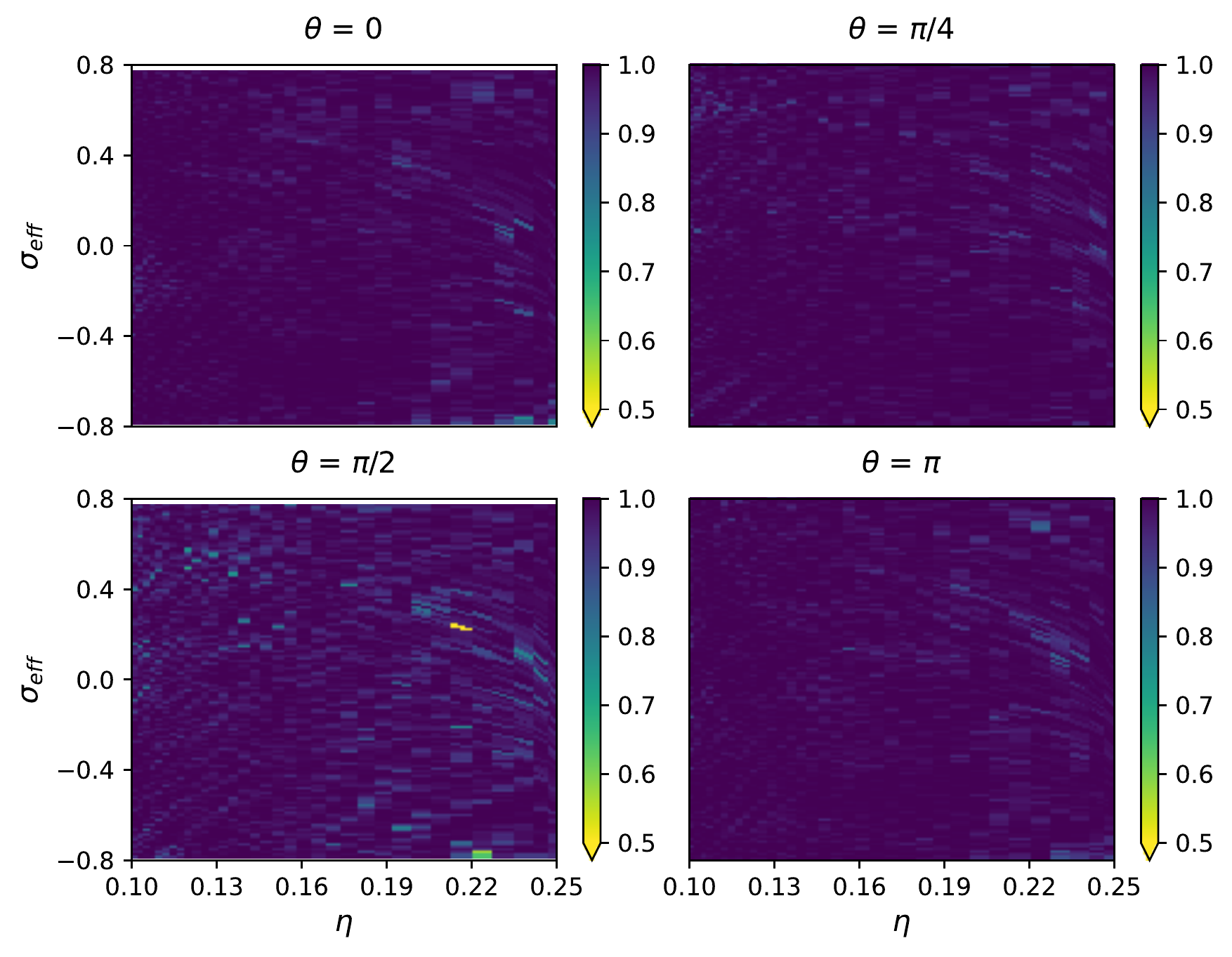}
} 
\caption{\textbf{Overlap between ground truth signals and those predicted by deterministic AI models} Overlap between ground 
truth and AI-predicted signals 
in terms of symmetric mass-ratio, \(\eta\), and effective spin, \(\sigma_{\mathrm{eff}}\), for a sample of inclination angles.}
\label{fig:oe_vs_theta}
\end{figure}

\noindent Figure~\ref{fig:ae_vs_theta} summarizes the 
accuracy with which deterministic AI models estimate 
the parameters of higher order modes for all mass-ratios 
and spins for a sample of inclination angles. 
There we notice that 
predictions degrade in accuracy for edge-on 
systems, i.e., \(\theta=\pi/2\). Note that a mild
shift is seen in the predicted values of the inclination angle, \(\theta\), 
from its ground
truth value in panels with \(\theta =
\{0, \pi\}\), although the distribution has the true value well within a \(1-\sigma\) deviation from its median. 
This shift is likely because \(0\) and \(\pi\)
correspond to the domain boundary of \(\theta\) as
\(0\leq \theta < \pi\), which leads to a one-sided
wrap-around of the recovered 
distribution of \(\theta\) values, and in turn
leads to a shift in the median
recovered value toward the boundary. 
We provide a more comprehensive analysis of 
these results in Figure~\ref{fig:oe_vs_theta}, where we 
show overlap calculations between ground truth and predicted 
signals in terms of symmetric mass-ratio and effective 
spin \((\eta, \sigma_{\mathrm{eff}})\) for all mass-ratios 
and spins under consideration for a sample of 
inclination angles. Again, here we find that our 
results are optimal for all angles except for 
edge-on binaries. We can understand this if we recall that 
for \(\theta=\pi/2\) we lose half of the information 
we feed into our AI models since 
\(h_{\times}\left(t, \theta=\pi/2\right)\rightarrow 0\). 
In summary, our deterministic AI models provide 
informative point parameter estimation results 
for the parameter space under consideration. 

\subsection{Comparison to other machine learning methods}

\begin{figure}[h!]
\centerline{
\includegraphics[width=1.0\linewidth]{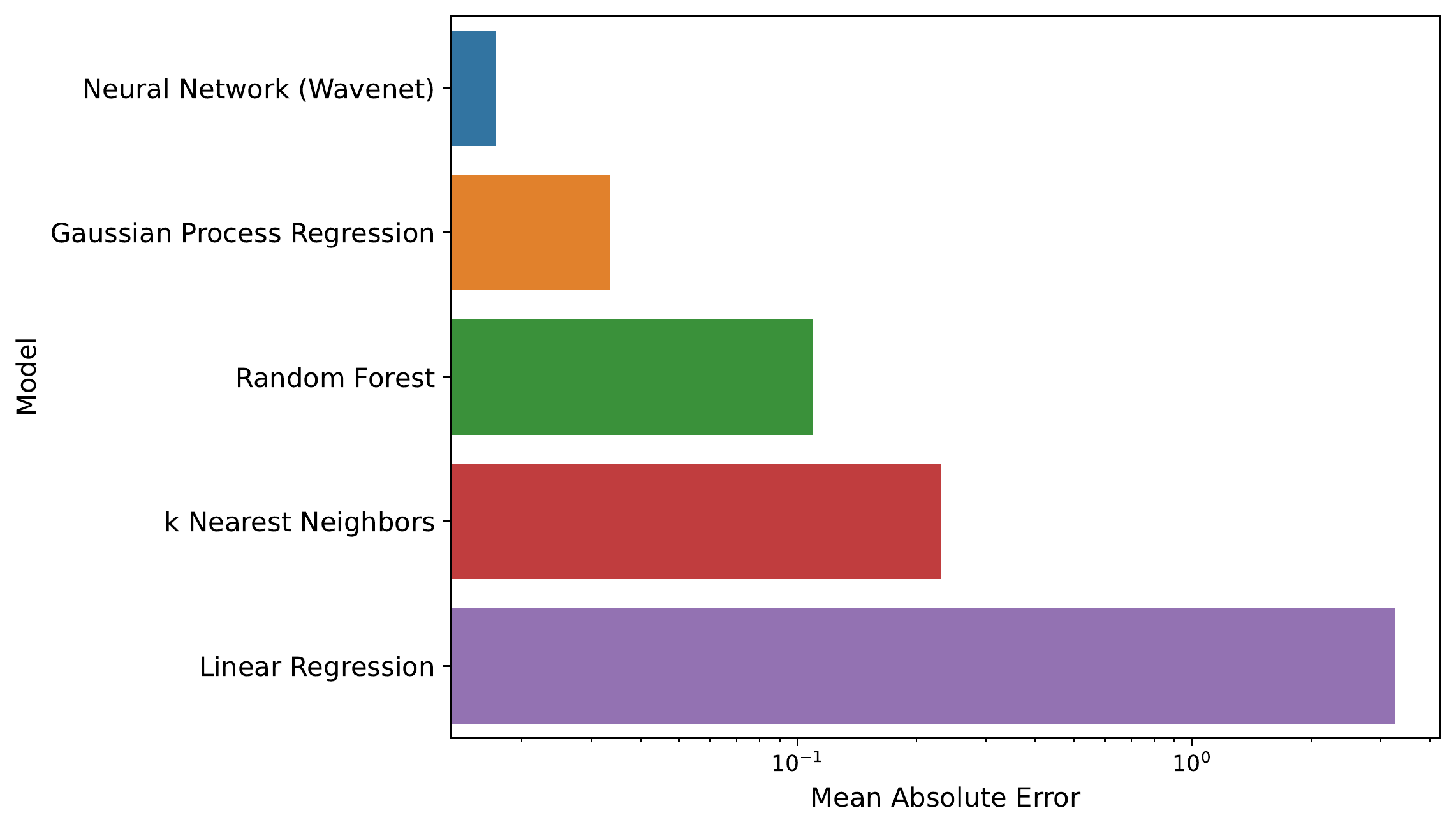}
} 
\caption{\textbf{Comparison between AI and traditional machine learning methods}
AI predictions are compared with machine learning methods considering 
a reduced dataset with mass-ratios \(q\in\{4.0, 4.1\}\) and 
inclination angles \(\theta\in\{0.0, 0.1\}\), and spin combinations. This 
approach is needed since these methods are sub-optimal to handle tens 
of millions of waveforms for training purposes. The trained 
machine learning models were then tested on a dataset with \(q=4.05\), \(\theta=0.05\) and all spin combinations. The mean absolute errors 
shown are the average of absolute errors over all four parameters 
\((q, s_1^z, s_2^z, \theta)\).}
\label{fig:ai_vs_t_ml}
\end{figure}

We have compared the predictions of our AI models with other 
machine learning algorithms, including, Gaussian Process Regression, 
Random Forest, k-Nearest Neighbors and Linear Regression. The results 
of this analysis are shown shown in Figure~\ref{fig:ai_vs_t_ml}.

For this comparison, we had to reduce the amount of training data, since 
traditional machine learning models are sub-optimal to handle the 
TB-size training datasets used to train our AI models. Thus, we 
considered a reduced data set that describes 
black hole binaries with mass-ratios \(q\in\{4.0, 4.1\}\) 
and inclination angles \(\theta\in\{0.0, 0.1\}\), and all 
possible spin combinations. Then, we tested these machine 
learning models on a test set with \(q=4.05\) and \(\theta=0.05\), 
and all possible spin combinations. Using the same metric we have 
considered above to quantify the performance of our AI models, 
e.g., mean absolute error, Figure~\ref{fig:ai_vs_t_ml} presents the 
average absolute errors over all four parameters under 
consideration, \((q, s_1^z, s_2^z, \theta)\). It is apparent that even 
in this scenario that provides a clear advantage to traditional 
machine learning models, neural networks still provide better results.

\noindent These results demonstrate the ability of AI 
to search across the signal manifold of higher order 
modes, and pinpoint a set of parameters, 
\(\{q, s_1^z, s_2^z,\theta\}\), that best describes 
the properties of waveforms that include higher-order modes. 
While this is informative, we also want to know the 
uncertainty associated with such AI-predicted 
values. To extract such information, we estimate 
posterior distributions using a combination of 
\texttt{WaveNet} with normalizing flow, as described 
in Section~\ref{sec:method}. 

\subsection{Probabilistic AI models}

\noindent We have selected binary black hole systems 
to quantify the ability of AI to reconstruct the 
astrophysical parameters of systems that are known to 
be hard to characterize. In particular, we consider 
comparable mass-ratio binaries, for which it is difficult 
to tell apart individual spins, as well as asymmetric 
mass-ratio systems, for which it is difficult to accurately 
constrain the spin of the secondary. 

We directly compare our AI-driven results with the 
Bayesian inference 
\texttt{PyCBC Inference} toolkit~\cite{Biwer:2018osg}. 
For consistency with AI-driven analysis, the inference results 
produced by \texttt{PyCBC Inference} assume 
noiseless signals, and a flat power spectral density. 
We scale the dimensionless signals used
above to the source location of a fiducial real event, 
GW150914, and use a flat noise power spectral density 
with amplitude set to the median noise level 
between $20 - 2048$Hz of the zero-detuning high-power 
design sensitivity curve for LIGO instruments~\cite{ZDHP:2018}.

\begin{figure}[h!]
\centering
\begin{subfigure}{\textwidth}
\raisebox{9.9cm}{
\hspace*{8.5cm}\includegraphics[width=.45\linewidth]{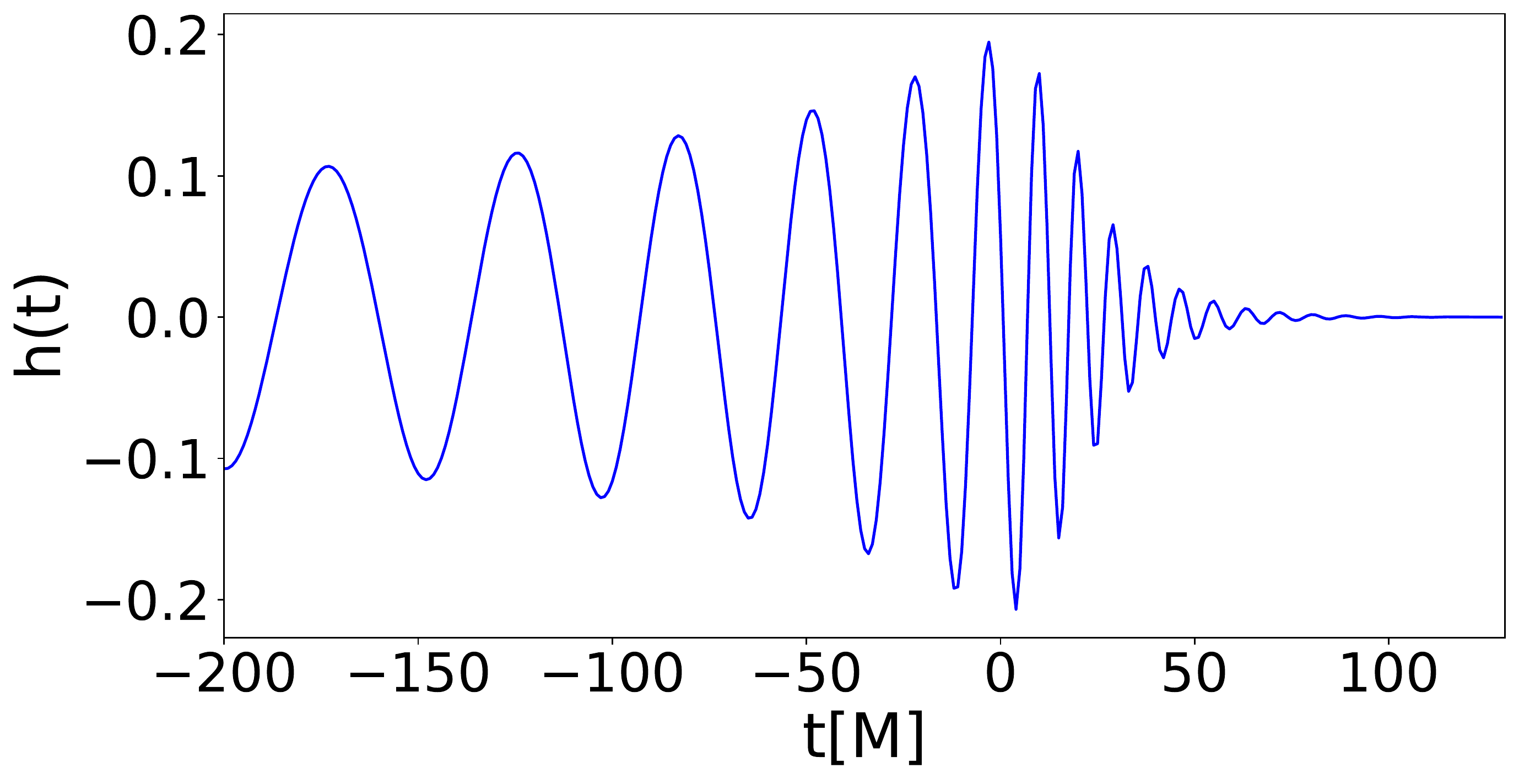}}\hspace*{-16.6cm} 
\raisebox{0cm}{
\includegraphics[width=\linewidth]{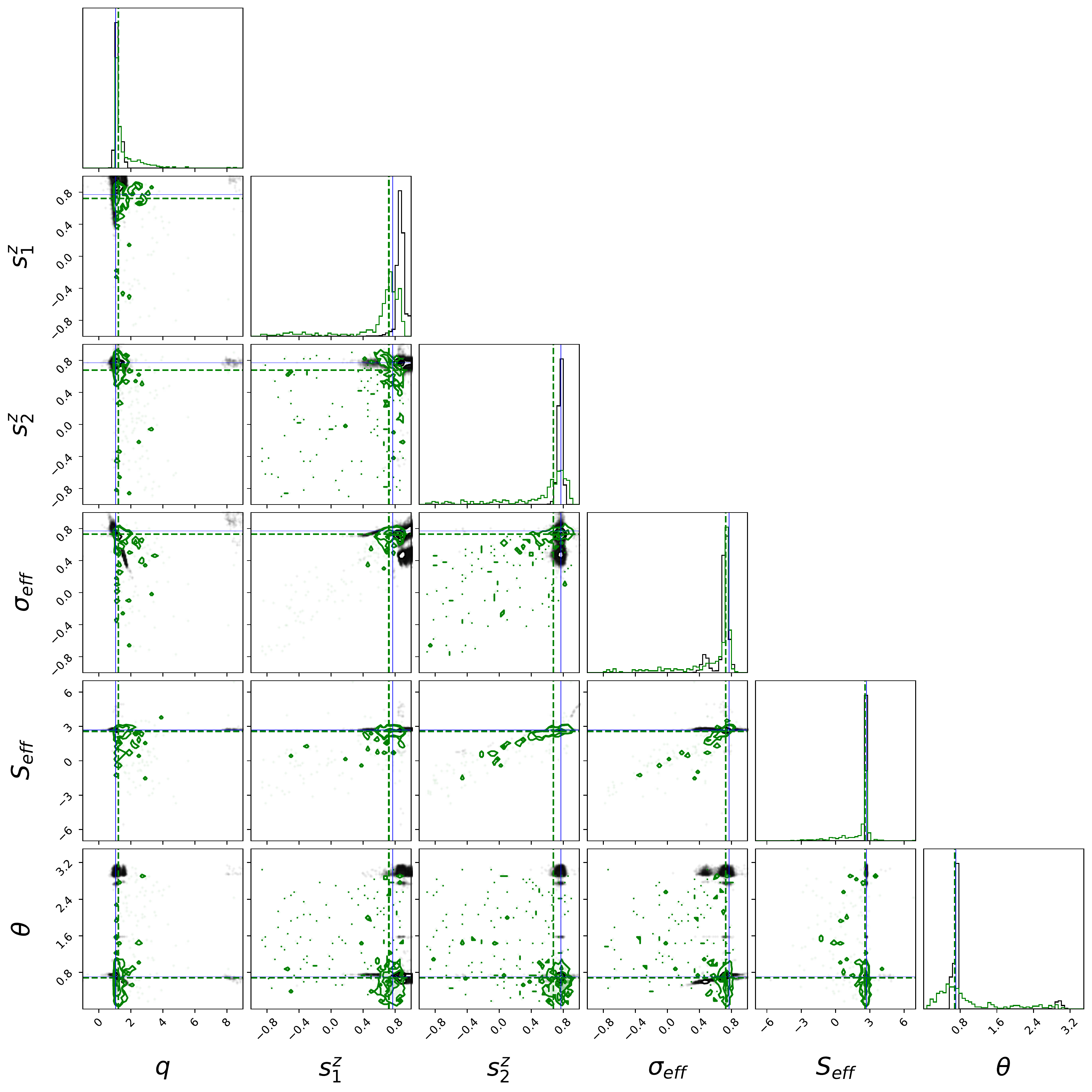}}
\end{subfigure}
\caption{\textbf{Probabilistic AI and Bayesian inference results, $q=1$ case.} AI posterior distributions (in black), \texttt{PyCBC Inference} results (in green), and ground truth values (in blue) for an equal mass-ratio binary black hole merger. AI histograms show the distribution of $100,000$ samples drawn from the posterior.}
\label{fig:nflow_23998}
\end{figure}

\begin{figure}[h!]
\centering
\begin{subfigure}{\textwidth}
\raisebox{9.9cm}{
\hspace*{8.5cm}\includegraphics[width=.45\linewidth]{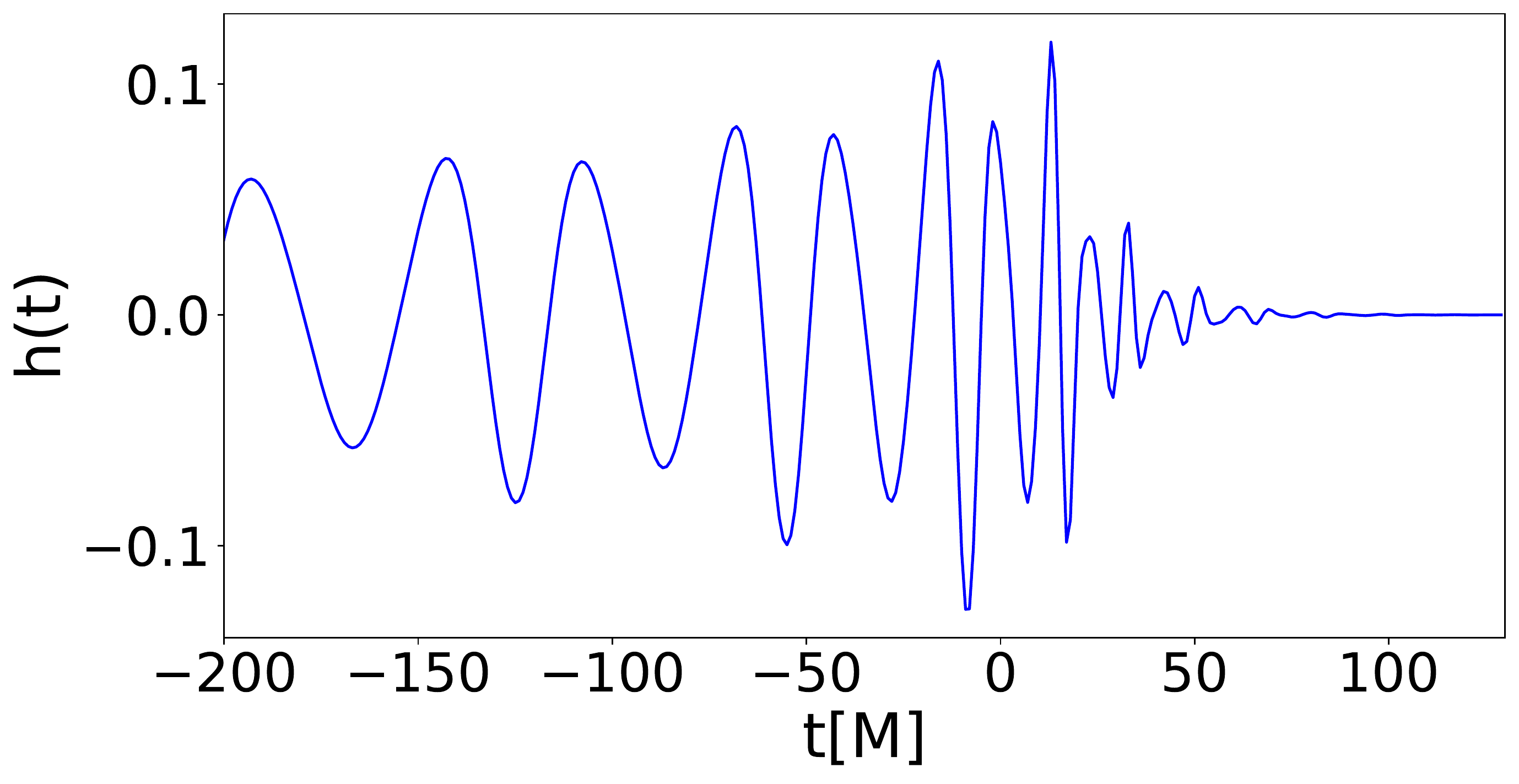}}\hspace*{-16.6cm} 
\raisebox{0cm}{
\includegraphics[width=\linewidth]{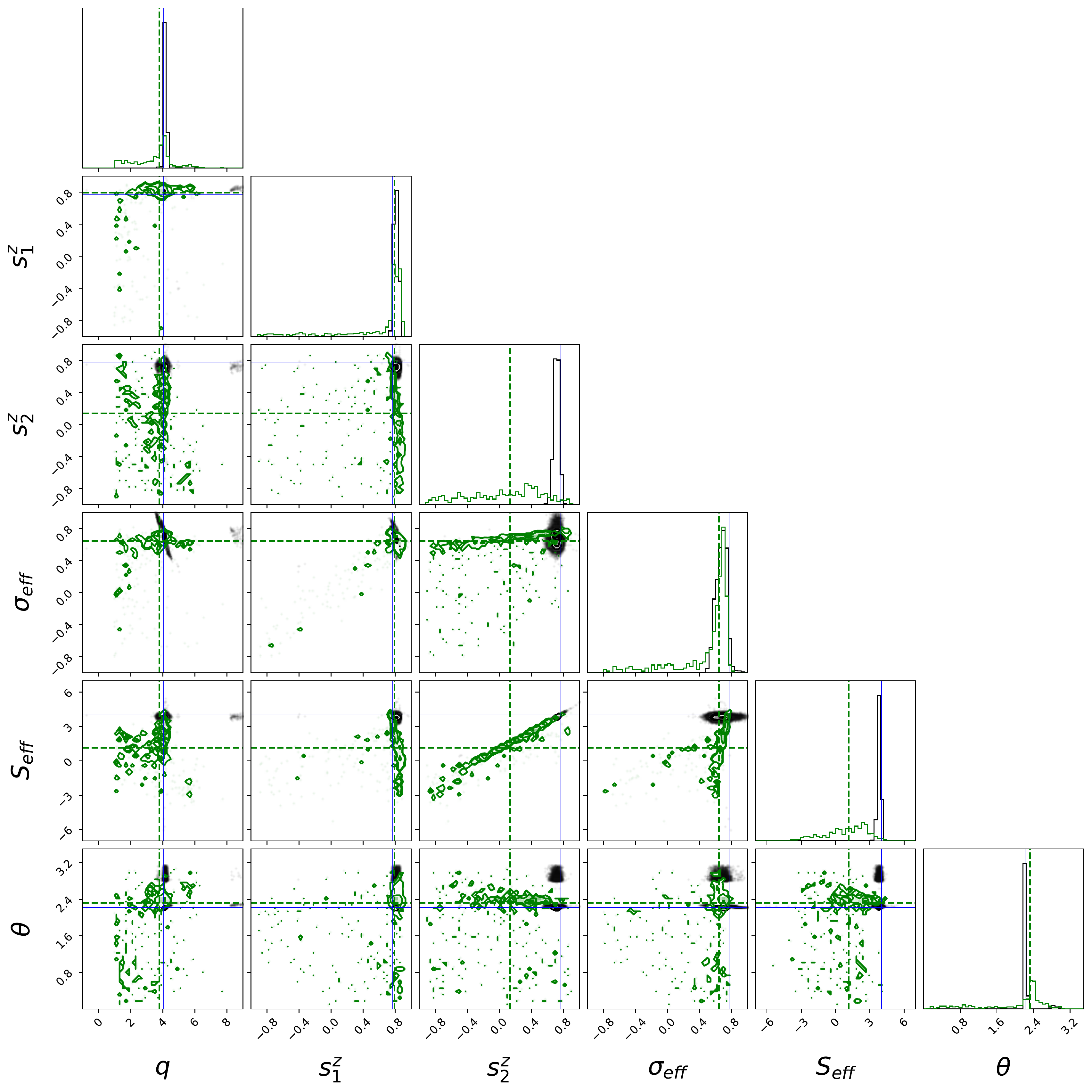}}
\end{subfigure}
\caption{\textbf{Probabilistic AI and Bayesian inference results, $q=4$, positive and aligned spin case.} As Figure~\ref{fig:nflow_23998}, but now for a binary with 
mass-ratio \(q=4\) whose binary components are rapidly spinning.}
\label{fig:nflow_383995}
\end{figure}

\begin{figure}[h!]
\centering
\begin{subfigure}{\textwidth}
\raisebox{9.9cm}{
\hspace*{8.5cm}\includegraphics[width=.45\linewidth]{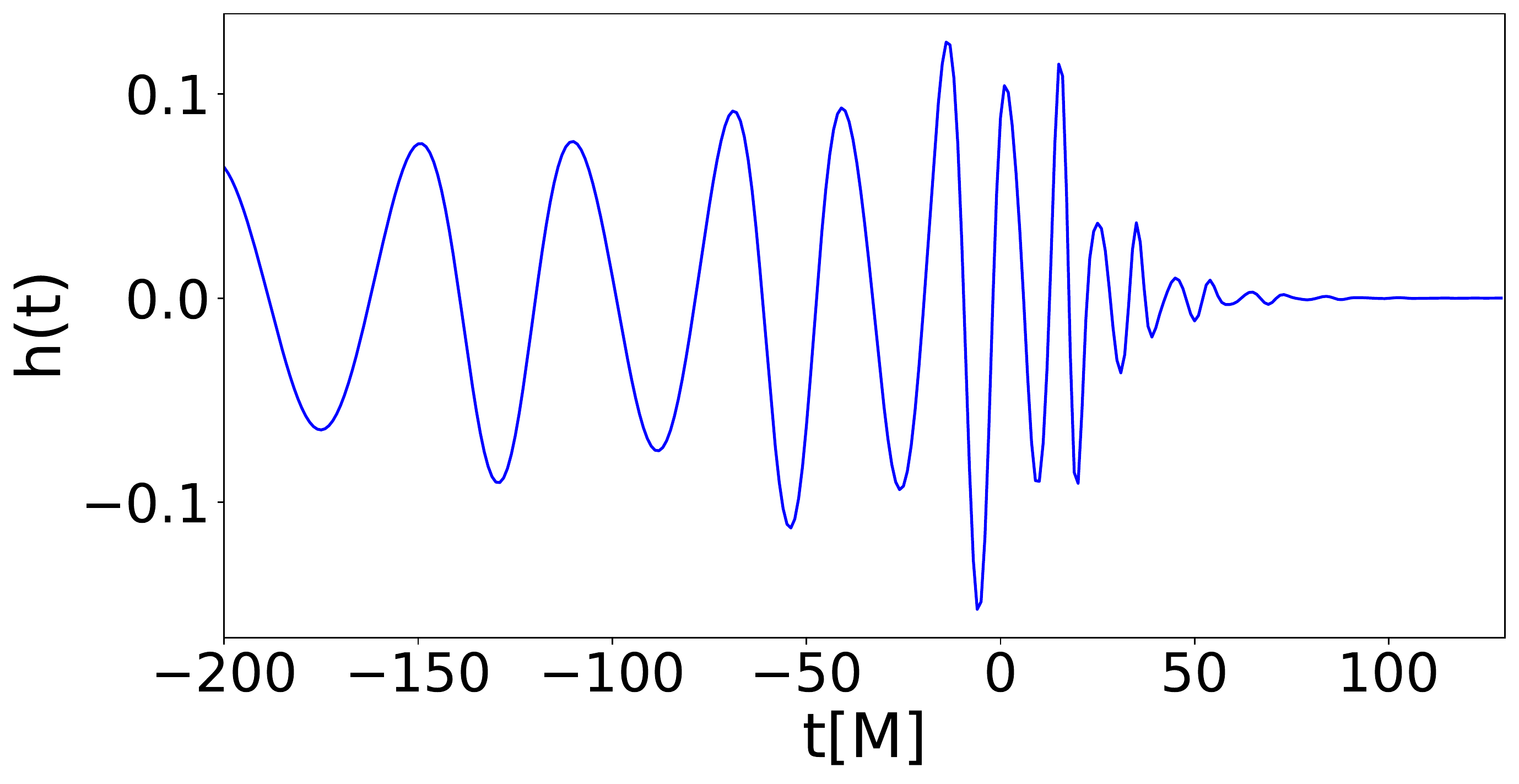}}\hspace*{-16.6cm} 
\raisebox{0cm}{
\includegraphics[width=\linewidth]{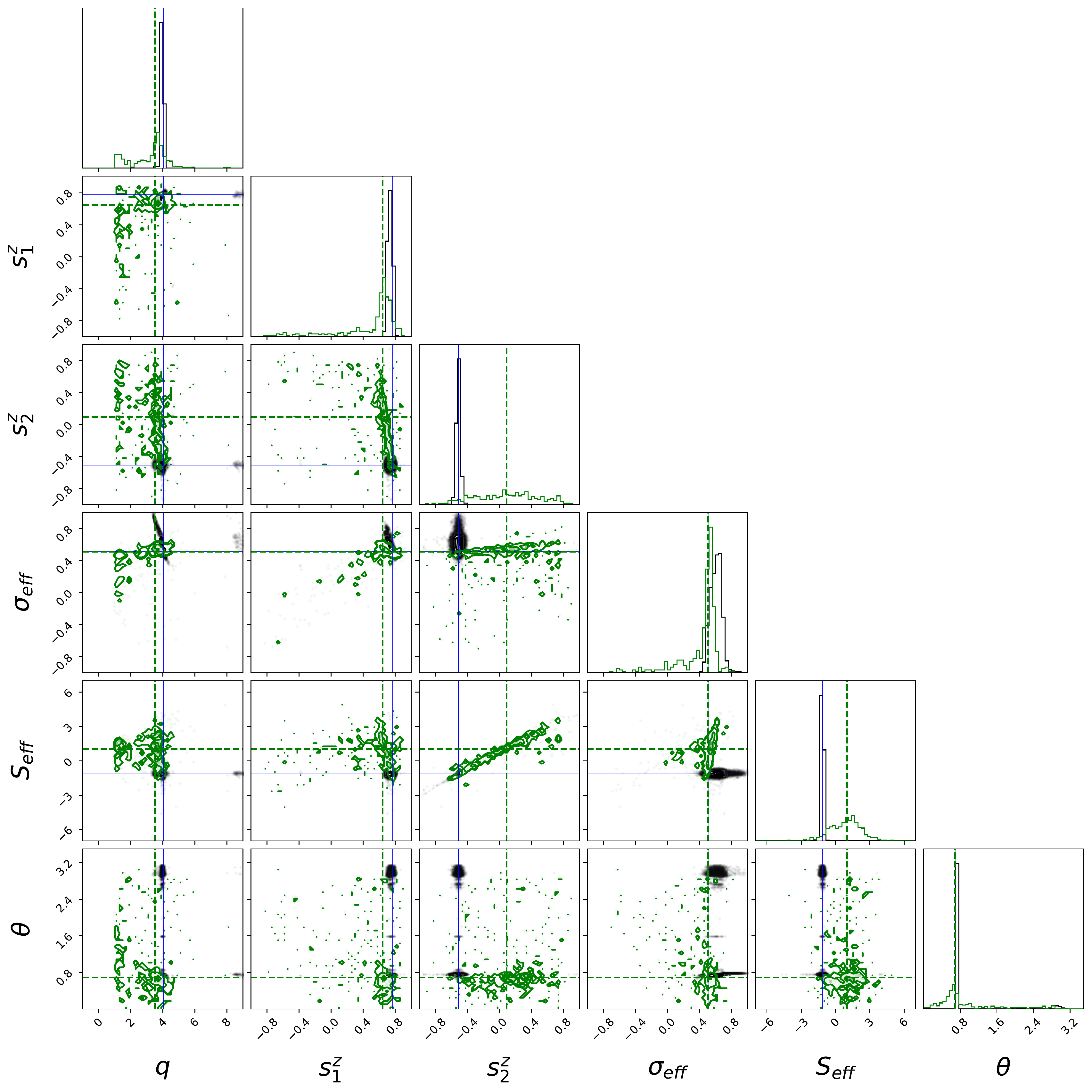}}
\end{subfigure}
\caption{\textbf{Probabilistic AI and Bayesian inference results, $q=4$, anti-aligned spin case.} As Figure~\ref{fig:nflow_383995}, but now the secondary is 
slowly rotating in the \(-z\) direction.}
\label{fig:nflow_383508}
\end{figure}

\begin{figure}[h!]
\centering
\begin{subfigure}{\textwidth}
\raisebox{9.9cm}{
\hspace*{8.5cm}\includegraphics[width=.45\linewidth]{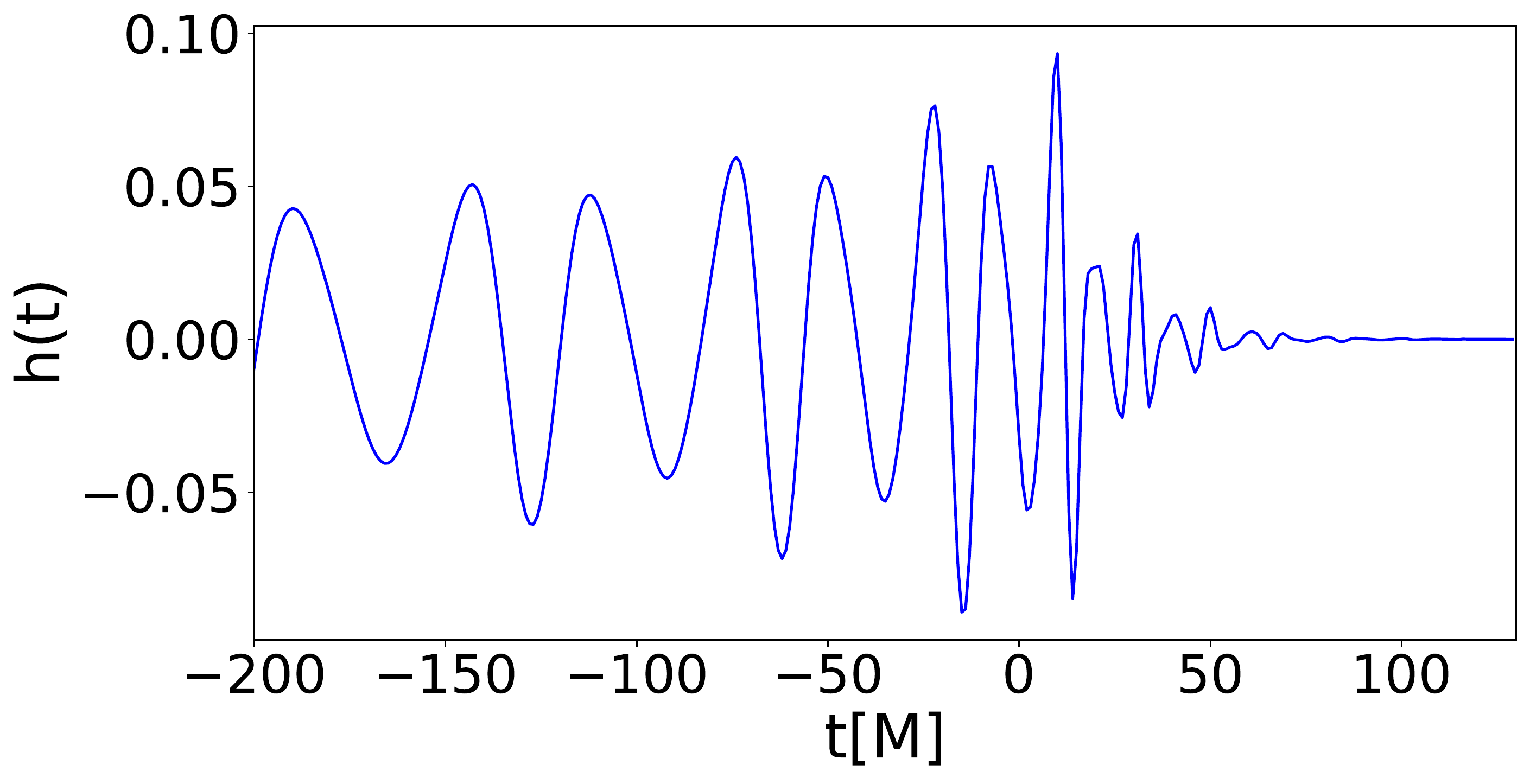}}\hspace*{-16.6cm} 
\raisebox{0cm}{
\includegraphics[width=\linewidth]{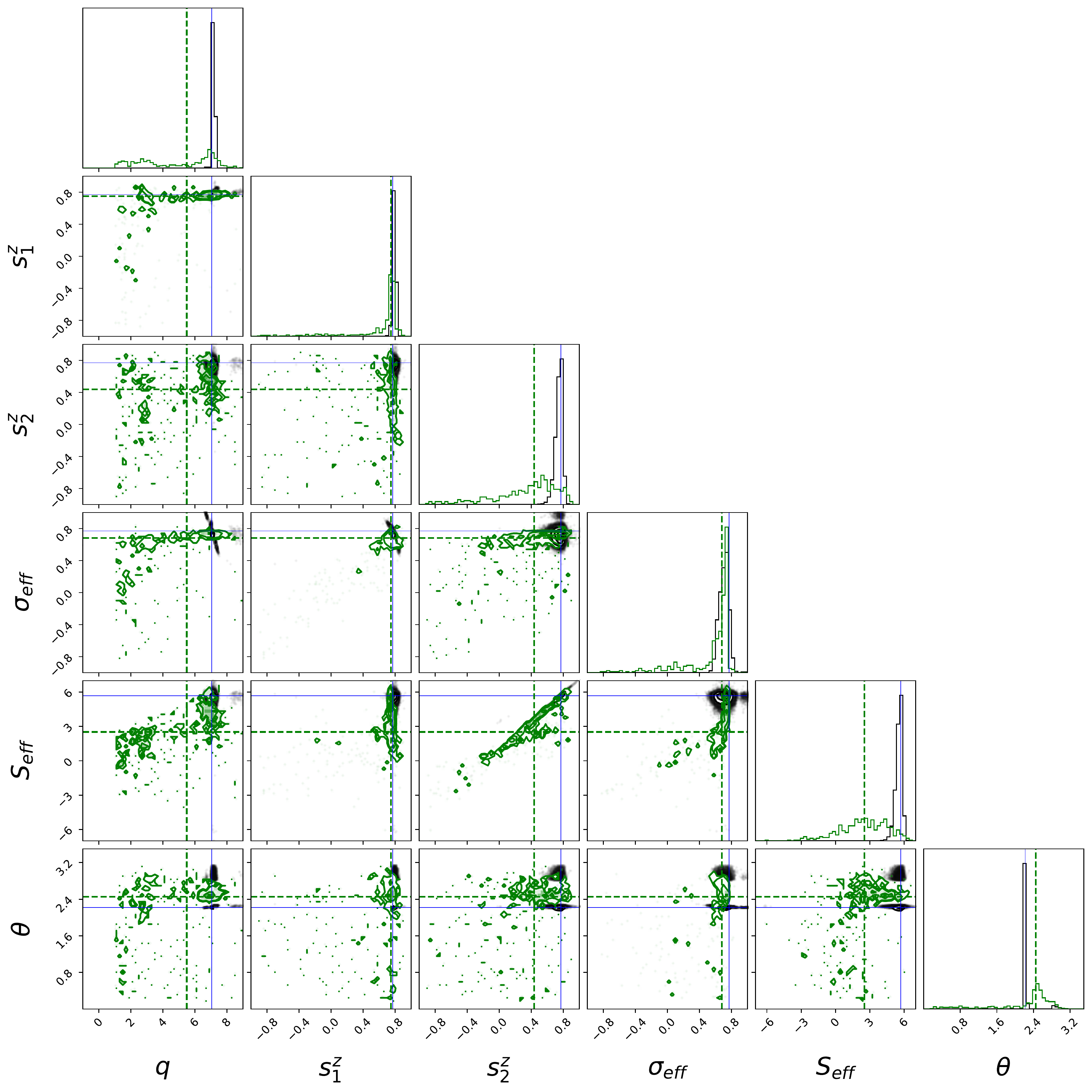}}
\end{subfigure}
\caption{\textbf{Probabilistic AI and Bayesian inference results, $q=7$, positive and aligned case.} As Figure~\ref{fig:nflow_23998}, 
but now for a binary with 
mass-ratio \(q=7\) whose binary components are rapidly spinning.}
\label{fig:nflow_743995}
\end{figure}

\begin{figure}[h!]
\centering
\begin{subfigure}{\textwidth}
\raisebox{9.9cm}{
\hspace*{8.5cm}\includegraphics[width=.45\linewidth]{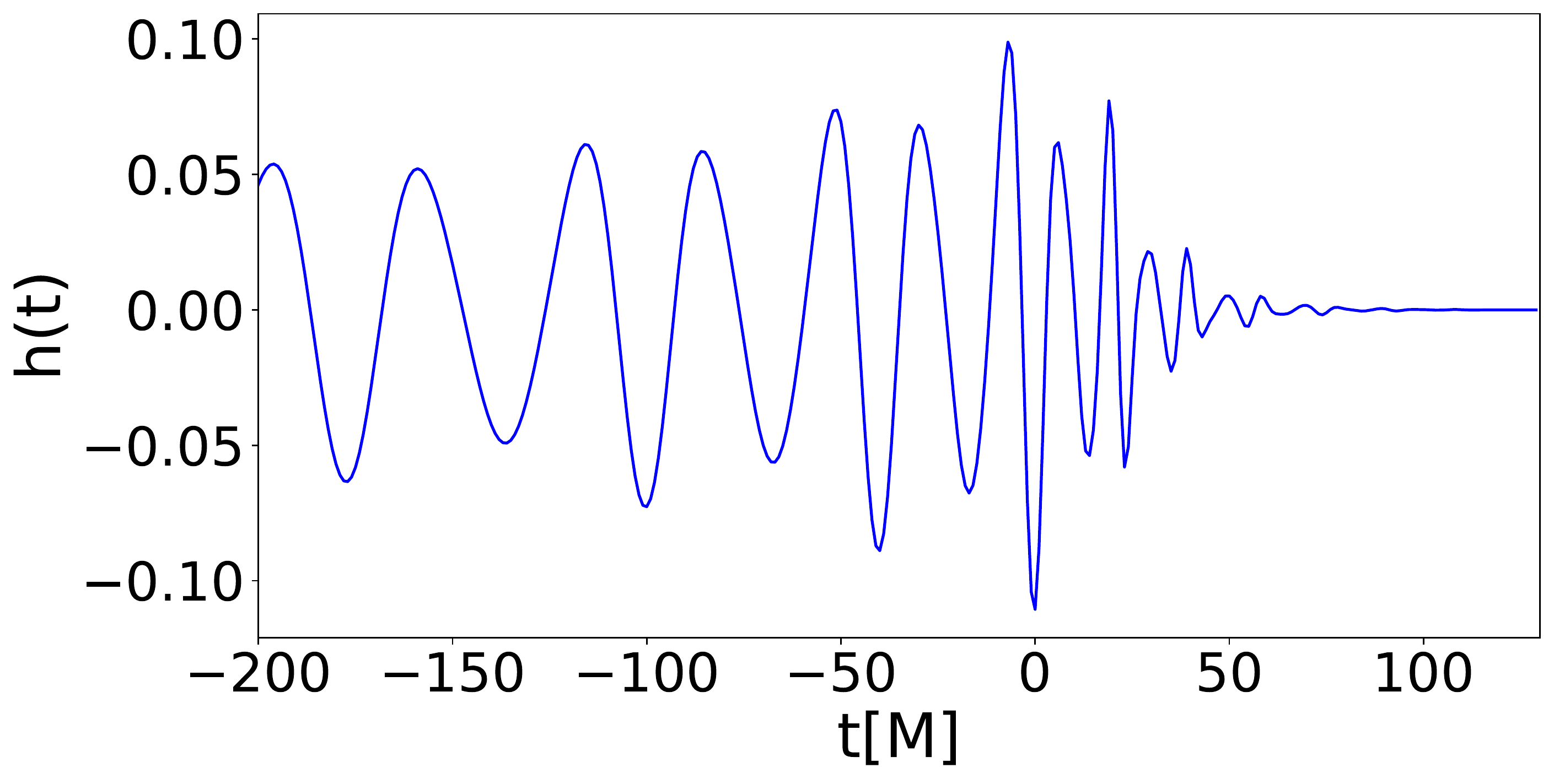}}\hspace*{-16.6cm} 
\raisebox{0cm}{
\includegraphics[width=\linewidth]{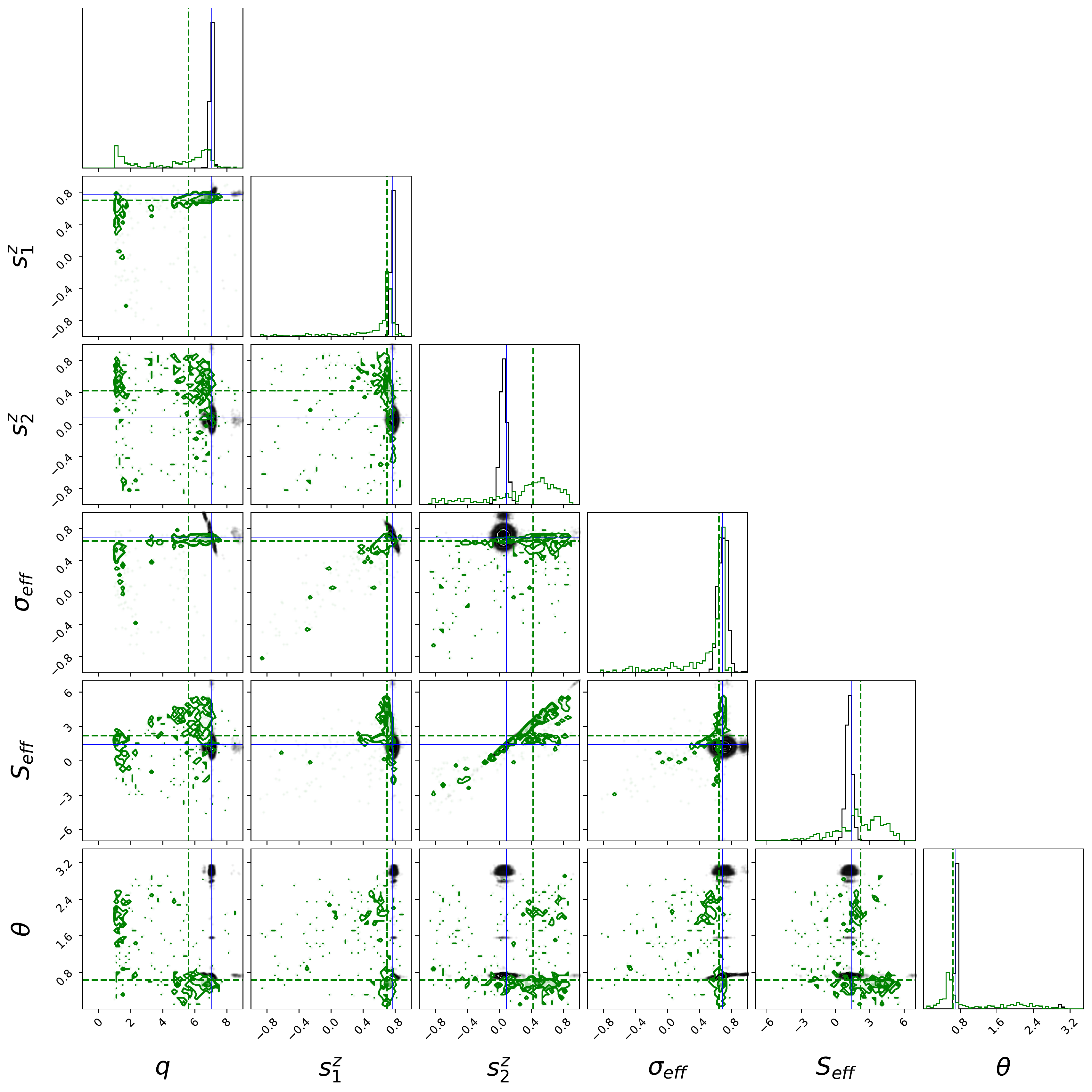}}
\end{subfigure}
\caption{\textbf{Probabilistic AI and Bayesian inference results, $q=7$, with primary rapidly rotating and secondary nearly spinless.} As Figure~\ref{fig:nflow_743995}, but now for a \(q=7\) binary 
with a rapidly rotating primary and a nearly non-spinning 
secondary.}
\label{fig:nflow_743733}
\end{figure}

Below we present probabilistic 
parameter estimation results for six 
astrophysical parameters: 
\((q, s^z_1, s^z_2, \sigma_{\textrm{eff}}, 
S_{\textrm{eff}}, \theta)\), using the following 
nomenclature: ground truth values are 
shown in blue; AI results are shown in black; 
\texttt{PyCBC Inference} results are shown in green:

\begin{compactitem}
\item Figure~\ref{fig:nflow_23998} presents results for an 
equal mass binary black hole merger. These results show that 
our AI model produces sharp, narrow distributions that 
provide informative constraints for the astrophysical 
parameters that describe this signal. We also notice 
that \texttt{PyCBC Inference} provides constraints that 
agree with ground truth values, though these distributions 
tend to be broad and with long tails, in particular 
for the spin of the secondary, \(s^z_2\), and 
the inclination angle, \(\theta\). 
\item Figures~\ref{fig:nflow_383995} 
and~\ref{fig:nflow_383508} present  
\(q=4\) binaries whose i) primary and secondary are  
rapidly rotating and aligned, and ii) primary is 
rapidly rotating and the secondary is moderately rotating 
and anti-aligned, respectively. Here again, AI inference 
results and ground truth 
values are in close agreement. \texttt{PyCBC Inference} 
results provide broader distributions for all 
parameters of interest, or uninformative constraints, 
in particular for the spin of the secondary. It is 
also noteworthy that while \texttt{PyCBC Inference} does 
not provide informative constraints for the spin of the 
secondary, estimates for the effective spin parameter, 
\(\sigma_{\textrm{eff}}\), are indeed informative, 
though they still present long tail distributions. 
\item Figures~\ref{fig:nflow_743995} and~\ref{fig:nflow_743733} 
present results for \(q=7\) binaries for two 
configurations. First, both binary components are 
rapidly rotating and aligned. Second, the primary 
is rapidly rotating and the secondary is nearly spinless. 
We selected these two configurations to illustrate that 
in either scenario AI estimates do provide narrow distributions, 
and informative constraints for all parameters under 
consideration, and in particular for the spin of the 
secondary. We now notice that for asymmetric 
mass-ratio black hole mergers \texttt{PyCBC Inference} 
does not provide tight constraints for the mass-ratio. 
Furthermore, this traditional Bayesian approach does 
not provide informative constraints for the secondary of 
the binary, and the inclination angle has a broad, long-tailed 
distribution. As before, the effective 
spin parameter, \(\sigma_{\textrm{eff}}\), provides informative 
constraints, though with a long tail.
\end{compactitem}


\noindent \paragraph{Benchmark results} These results 
provide evidence that AI surrogates are capable of 
learning and inferring the physics that describes  
quasi-circular, spinning, non-precessing, higher-order 
waveform modes of binary black hole mergers. In addition 
to these findings, we also provide results to compare the 
computational efficiency of AI and traditional Bayesian 
inference to produce Figures~\ref{fig:nflow_23998} 
to~\ref{fig:nflow_743733}. In terms of computational performance 
we found that

\begin{cititemize2}
\item We produced AI results presented in these figures by drawing 
100,000 samples from the posterior distribution using 
normalizing flow. This is done in less than a second using 
a single A100 NVIDIA GPU for all cases under consideration, 
i.e., irrespective of the properties of the binary black 
hole merger. 
\item For \texttt{PyCBC Inference}, we used 10 single-thread 
processes on an AMD EPYC 7352 with 24 physical cores to 
draw a similar amount of samples from the posterior distribution 
within 
\subitem 1 hour 41 minutes for Figure~\ref{fig:nflow_23998}
\subitem 3 hours 49 minutes for Figure~\ref{fig:nflow_383995}
\subitem 4 hours 21 minutes for Figure~\ref{fig:nflow_383508}
\subitem 6 hours 37 minutes for Figure~\ref{fig:nflow_743995}
\subitem 2 hours 20 minutes for Figure~\ref{fig:nflow_743733}
\end{cititemize2}

\noindent In summary, our probabilistic AI surrogates 
are between three to four orders of magnitude faster than 
traditional Bayesian inference methods. These results also 
indicate that we should clearly 
differentiate the physics we can infer from 
gravitational waves, and how the choice of 
signal processing tools will enhance or limit the science 
reach of our studies. For instance,  it has been 
argued in the literature that 
it would be difficult to infer the spin of the secondary 
for asymmetric mass-ratio binaries, since the rotation of 
the lighter black hole has a marginal influence on the 
morphology of the waveform. What we are learning from this 
study is that we ought to tell apart the physics of the 
problem from the signal processing tools utilized to 
study the astrophysical properties of compact binaries, 
and for that matter for any system. 
It is true that traditional Bayesian inference 
is unable to provide informative constraints for the 
spin of the secondary, even for moderately asymmetric 
mass-ratio mergers, as we see in Figures~\ref{fig:nflow_383995} 
and~\ref{fig:nflow_383508}; or for asymmetric mass-ratio 
systems, as we see in 
Figures~\ref{fig:nflow_743995} and~\ref{fig:nflow_743733}.
However, our probabilistic AI models 
do provide informative 
constraints for the spin of the primary and secondary 
for binaries with mass-ratios \(1\leq q \leq 8\). 
This is because the subtle features and patterns that the spin of 
the secondary imprints in the waveform signal are identified 
and learned by AI, and this empowers us to infer 
these astrophysical parameters from complex signals that 
include higher-order waveform modes.

We also learn another piece of information from this 
analysis. As the broader community continues to develop AI-driven 
methodologies for accelerated inference, we should endeavor 
to develop novel AI tools, and to not limit the capabilities 
of these algorithms to simply accelerate parameter estimation 
analyses, providing nearly identical results to traditional 
Bayesian inference pipelines. Doing so will limit our ability 
to learn new insights from gravitational wave observations. 
As we have learned from this study, AI methods that 
are designed to mimic Bayesian pipelines 
will provide uninformative constraints on the 
spin distribution of compact binaries, in particular 
for the secondary, thereby limiting the knowledge or 
insights we could gain about the formation channels 
of these sources. Similarly, uninformative 
constraints for the inclination angle 
of these sources would have implications for 
gravitational wave cosmology.

This study has shown that in the case where 
we look at the signal manifold of higher-order 
waveform signals for black hole mergers, there is a 
wealth of astrophysical information we can 
extract from these signals using probabilistic 
AI methods. We have also learned 
that Bayesian approaches cannot capture 
features and patterns that enable the measurement 
of important astrophysical parameters, and that 
this is not a result of biases introduced by 
noise, since we are not considering the effect of noise 
at this stage. These signal processing limitations 
in parameter estimation are inherent to traditional 
Bayesian inference.

Another important result result of this paper is that 
we have designed a methodology to train AI models 
that adequately handle training datasets that include tens of 
millions of modeled waveforms, thereby paving the 
way to extend this analysis for the case 
in which these types of signals are contaminated with 
simulated and advanced LIGO noise. The methods introduced 
in this paper will enable us to quantify the biases 
introduced by noise in parameter estimation analyses, 
and how to handle them to extract 
informative AI-driven parameter estimation results using 
higher order gravitational wave modes.

\section{Conclusions}
\label{sec:end}

\noindent We have developed scalable and computationally 
efficient methods to design AI models that are 
capable of characterizing the signal manifold of 
higher order wave modes of quasi-circular, spinning, 
non-precessing binary black hole mergers. Our approached 
enabled us to train several AI models using a dataset 
of over 14 million waveforms within 3.4 hours with 256 nodes, 
equivalent to 1,536 NVIDIA V100 GPUs, achieving 
optimal convergence and state-of-the-art regression 
results.

We have demonstrated that AI can abstract knowledge from 
time-series data that help constrain the physical 
parameters that determine the 
dynamical evolution of higher order modes of black 
hole mergers. In particular, we have presented evidence 
that AI provides deterministic and probabilistic predictions 
that tightly constrain the mass-ratio, individual spins, 
inclination angle, and effective spin parameters for 
a variety of astrophysical scenarios. We also 
found that deterministic and probabilistic AI predictions 
are consistent with each other, and in good accord with 
ground truth physical parameters.

We have also demonstrated that our AI surrogates 
outperform other machine learning methods (encompassing 
Gaussian regression, random forest, k-nearest neighbors, and 
linear regression), and 
\texttt{PyCBC Inference} both in terms of 
computational efficiency and accuracy. 
The results we have introduced in this article provide 
benchmarks for the expected performance of AI to estimate the 
astrophysical parameters of binary black hole mergers in the 
absence of noise. In future work, we will present studies 
for the impact of simulated and advanced LIGO noise to 
conduct informative AI-driven inference for high dimensional 
waveform manifolds.

\section{Acknowledgements}
\label{ack}
\noindent A.K. and E.A.H. gratefully acknowledge National 
Science Foundation (NSF) awards OAC-1931561 and 
OAC-1934757, and the Innovative and 
Novel Computational Impact on Theory and Experiment 
project `Multi-Messenger Astrophysics 
at Extreme Scale in Summit'. P.K. acknowledges support of
the Department of Atomic Energy, Government of India,
under project no. RTI4001, and by the Ashok and
Gita Vaish Early Career Faculty Fellowship at the
International Centre for Theoretical Sciences.
This material is based upon work supported 
by Laboratory Directed Research and Development (LDRD) 
funding from Argonne National Laboratory, 
provided by the Director, Office of Science, of the 
U.S. Department of Energy under 
Contract No. DE-AC02-06CH11357. 
This research used resources of the Argonne 
Leadership Computing Facility, which is a DOE Office of 
Science User Facility supported under Contract 
DE-AC02-06CH11357.
This research used 
resources of the Oak Ridge Leadership Computing Facility, 
which is a DOE Office of Science User Facility 
supported under contract no. DE-AC05-00OR22725. 
This work utilized resources supported 
by the NSF's Major Research Instrumentation program, 
the HAL cluster (grant no. OAC-1725729), 
as well as the University of Illinois at 
Urbana-Champaign. We thank \texttt{NVIDIA} for 
their continued support.

\bibliographystyle{ieeetr}
\spacingset{1}
\bibliography{book_references}

\end{document}